%% file: include.tex
\documentclass[
  aps,
  prl,
  amsmath,amssymb,
  superscriptaddress,
  reprint,
  twocolumn,
  nolongbibliography,
  secnumbers,
  footinbib
]{revtex4-2}

\usepackage{graphicx}
\usepackage[sectionbib]{chapterbib}
\usepackage{dcolumn}
\usepackage{bm}
\usepackage{physics}
\usepackage{siunitx}
\usepackage{hhline}
\usepackage{url}
\usepackage{caption}
\usepackage{subcaption}
\usepackage{mathtools}
\usepackage{amsmath}
\usepackage{amssymb}

\usepackage{mhchem}
\usepackage[dvipsnames]{xcolor}
\usepackage[
  breaklinks=true,
  colorlinks=true,
  linkcolor=Blue,
  citecolor = Blue,
  urlcolor = Blue,
  unicode
  ]{hyperref}

\usepackage[nameinlink]{cleveref}
\crefname{equation}{Eq.}{Eqs.}
\Crefname{equation}{Equation}{Equations}
\crefname{figure}{Fig.}{Figs.}
\crefname{section}{Appendix}{Appendix}
\Crefname{section}{Appendix}{Appendix}
\Crefname{figure}{Figure}{Figures}

\usepackage{caption}
\usepackage{enumerate}
\usepackage{enumitem}

\usepackage[T1]{fontenc}
\usepackage[utf8]{inputenc} 
\usepackage{tgtermes}
\usepackage{newtxmath}

\newcommand{\utokyo}{Department of Physics, The University of Tokyo, 7-3-1 Hongo, Bunkyo, Tokyo 113-0033, Japan}
\newcommand{\tokyotech}{Department of Electrical and Electronic Engineering, School of Engineering, Institute of Science Tokyo, 2-12-1 Ookayama, Meguro, Tokyo 152-8552, Japan}
\newcommand{\nims}{National Institute for Materials Science, 1-1 Namiki, Tsukuba, Ibaraki 305-0044, Japan}
\usepackage[normalem]{ulem}
\usepackage{quantikz}

\usepackage{color}

\newcommand{\hyphen}{\nobreakdash-\hspace{0pt}}

\newcommand{\pdvd}{\mathrm{p.v.}}
\newcommand{\Bzd}{B_z^{\, d}}

\begin{document}

\title{Visualization of Current-Driven Vortex Formation\\ in High-$T_c$ Cuprate Superconductors}
\author{Shunsuke Nishimura}
\affiliation{\utokyo}

\author{Takeyuki Tsuji}
\affiliation{\tokyotech}
\affiliation{%
 \nims
}%
\author{Takayuki Iwasaki}
\affiliation{\tokyotech}

\author{Mutsuko Hatano}
\affiliation{\tokyotech}

\author{Kento Sasaki}
\affiliation{\utokyo}

\author{Kensuke Kobayashi}
\affiliation{\utokyo}
\date{\today}

\input{letter_final.tex}

\input{include1.bbl}
\input{letter_SI_nightly.tex}

\input{include2.bbl}
\end{document}

%% file: letter_final.tex
\begin{abstract}
  Type-II superconductors exhibit hysteretic behavior due to the presence of quantum vortices, and the order in which temperature and external field are varied plays a decisive role. 
  Here we take current, rather than magnetic field, as the external drive. 
  We image the magnetic field of a high-$T_c$ cuprate superconductor strip after cooling.
  We confirm that even in zero magnetic field, current-biased cooling nucleates vortices within the strip. 
  With a small external magnetic field, the distribution is polarized opposite to the Lorentz-force direction. 
  These behaviors follow from the self-consistent relation between current and local field in steady flux flow. 
  Our findings show that current history is encoded as vortices. This reveals self-field effects that influence dc measurements and glassy transitions under drive.
\end{abstract}

\maketitle
\textit{Introduction}---Nonequilibrium phase transitions, especially in driven many-body systems, have recently emerged as broad topics across physics~\cite{Takeuchi2007,Takeuchi2010,Liu2025,Corte2008,Cates2015,Marjolein2019,Berthier2013,Vicsek1995,Bi2015,Bi2016,Zhou2013,Fitzpatrick2017,Sieberer2025,Henkel2008,Hinrichsen2006,Schmittmann1995,Zurek1985,Kibble1976,Bray1994}. These phenomena span liquid crystals~\cite{Takeuchi2007,Takeuchi2010,Liu2025}, colloids~\cite{Corte2008,Cates2015,Marjolein2019,Berthier2013}, active matter~\cite{Vicsek1995,Bi2015,Bi2016,Zhou2013}, and open quantum systems~\cite{Fitzpatrick2017,Sieberer2025}. 
In particular, superconducting quantum vortices provide a paradigmatic platform, as they undergo nonequilibrium phase transitions that share theoretical features with colloidal systems~\cite{Reichhardt2017,Mangan2008,Okuma2011}. 

In superconductors, vortices impose strong history dependence, and the sequence in which temperature and external field are varied is decisive. 
Regarding magnetic fields, the key distinction is between zero-field cooling (ZFC) and field cooling (FC). 
After ZFC, the superconductor first enters the Meissner state. 
When a field is applied afterward, a finite-sized sample exhibits vortex penetration at the edge, as described by the critical state model (CSM) which accounts for vortex entry/exit magnetic hysteresis~\cite{Bean1962,Bean1964,Jooss2002,Bending1999,Brandt1996,Tamegai1992,Grigorenko2001,Johansen1996,Blatter1994}. 
Under FC, the sample is cooled through $T_c$ in a constant field, then enters a mixed state with sample-wide flux trapping; the resulting patterns have been imaged directly~\cite{Kirtley2010,Fasano2008,Jooss2002,Kokubo2010,Brandt1996,Casola2018,Bending1999,Suderow2014,Bean1962,Zeldov1994Geometrical,Pearl1964,Kogan2003}.

Here we ask the current-driven analogue: what state emerges when a superconductor is cooled under a constant current rather than a magnetic field? 
When a current is applied \emph{starting from the Meissner state}, the CSM for transport current applies~\cite{Brandt1993,Zeldov1994Magnetization}, and edge-localized vortex entry has been confirmed experimentally~\cite{Gaevski1999}. 
By contrast, cooling across $T_c$ under fixed current instead of fixed field [current-biased cooling (CC)] has scarcely been examined, leaving room for experimental verification via magnetic imaging.

In CC, the nonlocal self-field generated by the current $J(x)$ plays a central role. We consider a superconducting strip in the thin-film limit, with thickness \(t_\mathrm{sc}\) and half-width \(W\) satisfying \(t_\mathrm{sc} \ll W\). The strip lies in the $xy$ plane, occupies \(|x|\le W\), and extends infinitely along \(y\). The sheet current,
$J(x)=\int_{-t_\mathrm{sc}/2}^{t_\mathrm{sc}/2} j(x,z)dz$, where \(j(x,z)\) denotes the volumetric current density, flows along the \(y\) direction and produces the perpendicular field component \(B_z(x)\) as~\cite{Brandt1993,Zeldov1994Magnetization}
\begin{equation}
  B[J](x)
  = \frac{\mu_0}{2\pi}\operatorname{p.v.}\int_{-W}^{W}\frac{J(u)}{u-x}du .
  \label{eq:biot-savart}
\end{equation}
For a uniform current $J(x)=J_0$, \cref{eq:biot-savart} reduces to~\cite{Zeldov1994Magnetization}
\begin{equation}
  B(x) = \frac{\mu_0 J_0}{2\pi}\log \left|\frac{W - x}{W + x}\right|.
  \label{eq:self-ampere}
\end{equation}
This self-field can nucleate vortices in CC. The current also propels them via the Lorentz force, inducing dissipative flux flow~\cite{Blatter1994,Bardeen1965,Kim1965,Brandt1995}.
As the temperature $T$ is lowered, especially in strong-pinning systems, vortex motion is progressively suppressed and eventually freezes~\cite{Blatter1994,Brandt1996,Koch1989,Yeshurun1996,Fisher1991}, yielding a crossover reminiscent of a dynamical-glass transition~\cite{Corte2008,Cates2015,Marjolein2019,Berthier2013,Reichhardt2017,Mangan2008,Okuma2011}.
Meanwhile, $J(x)$ and $B(x)$ are coupled by the Biot--Savart relation [\cref{eq:biot-savart}] and Faraday induction. Consequently, $J(x)$ can evolve self-consistently with vortex generation.

Hence, CC fundamentally differs from FC, which proceeds in a quasi-equilibrium manner; in CC, the system can imprint the vortex dynamics in the final vortex configuration. This scenario has not yet been experimentally verified.

In this Letter, we report the magnetic-field distribution of a \ce{YBa2Cu3O_{7-\delta}} (YBCO) thin-film strip during CC, observed using a quantum diamond microscope (QDM). We directly visualize vortex freezing under CC. From the magnetic-field maps, we reconstruct the current distribution. Contrary to a simple expectation, the distribution is not uniform across the strip; instead, the current density is enhanced at the strip center during vortex nucleation. We show that this behavior is consistent with steady-state, self-consistent solutions of flux-flow and flux-creep models. These results demonstrate that current history becomes imprinted as quantum vortices, and they provide new experimental guidance on the glassy crossover from flow to a frozen, disordered state out of equilibrium.

\begin{figure}[tbp]
  \begin{center}
    \includegraphics[width = 0.5\textwidth]{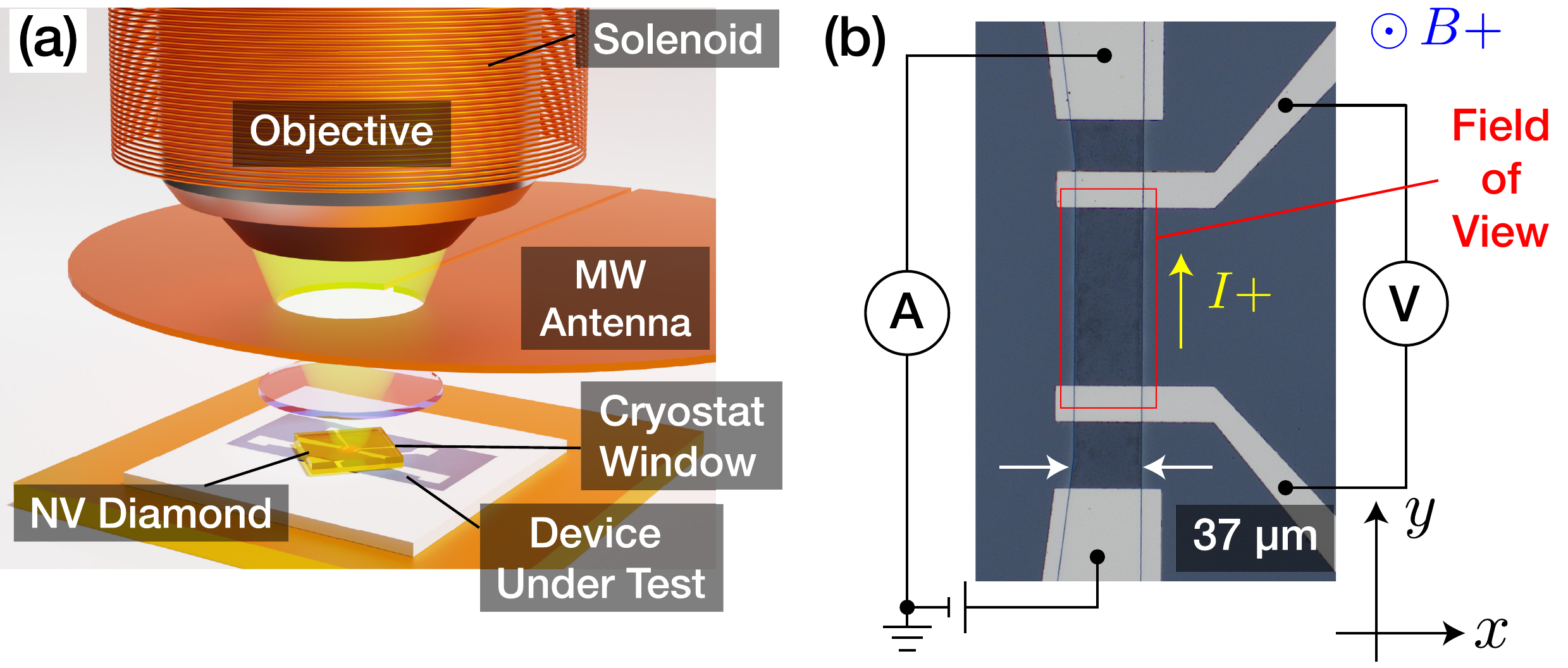}
    \caption{
      (a) Schematic of our QDM. 
      (b) Optical micrograph and circuit diagram of the YBCO strip. 
    }
    \label{fig:schematic}
  \end{center}
\end{figure}

\textit{Methods}---We use a QDM~\cite{Taylor2008,Levine2019} with a perfectly-aligned nitrogen-vacancy (NV) ensemble~\cite{Nishimura2023, Tsuji2022} of thickness \SI{2.4}{\micro\meter} grown on the diamond surface, which images the magnetic flux density perpendicular to the superconducting film~\cite{Nishimura2023}. 
\Cref{fig:schematic}(a) illustrates the experimental setup.
The sample is mounted on the stage of a cryostat with an optical window and temperature control.
NV-center photoluminescence is collected with an image sensor through the cryostat window. 
A YBCO film with $t_{\mathrm{sc}}=\SI{250}{\nano\meter}$ ($T_c=86.1$~K) is patterned into a strip of width $2W= \SI{37}{\micro\meter}$, as shown in \cref{fig:schematic}(b). 
A solenoid applies an out-of-plane static field $B_a$.
Additional details are provided in the Supplemental Information (SI)~\cite{SI}.

\begin{figure}[tbp]
  \begin{center}
    \includegraphics[width = 0.5\textwidth]{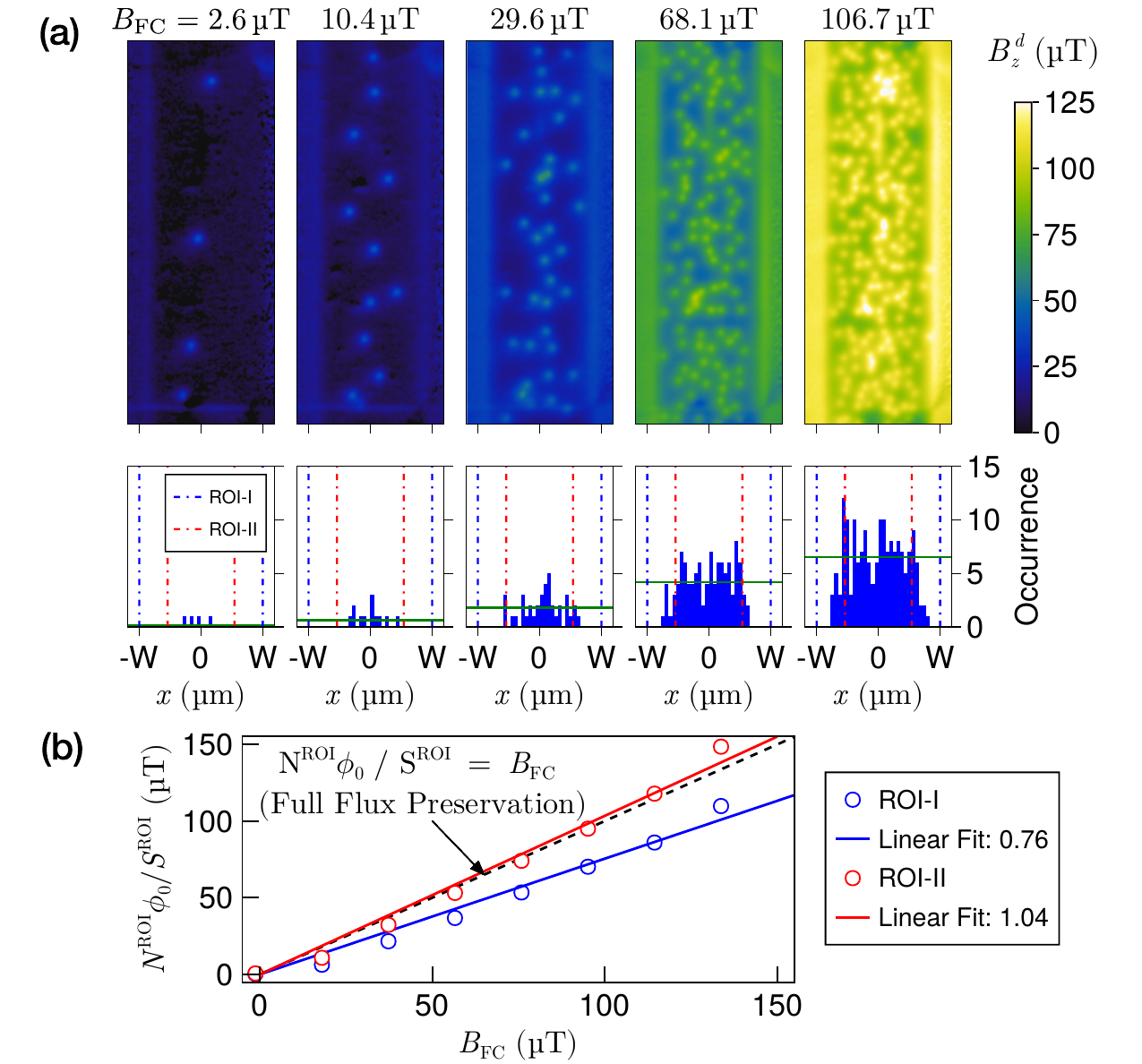}
    \caption{
      (a) (top) Magnetic-field maps obtained by FC at several $B_{\mathrm{FC}}$. 
      (bottom) Histograms of the vortex $x$-positions.
      (b) $N^{\mathrm{ROI}}\phi_0/S^{\mathrm{ROI}}$ as a function of $B_{\mathrm{FC}}$ (see text).
    }
    \label{fig:FC}
  \end{center}
\end{figure}

\textit{FC without current}---To validate vortex generation and its spatial distribution, we first perform FC under a magnetic field $B_{\mathrm{FC}}$ without current. Representative QDM maps of the out-of-plane magnetic field in the NV-layer plane, $B^{\, d}$, are shown in \cref{fig:FC}(a, top). Each isolated $\sim$\SI{90}{\micro\tesla} peak inside the strip corresponds to a single vortex. These data reveal a tendency for vortices to avoid the edges. 
The histogram of vortex $x$-positions [\cref{fig:FC}(a, bottom)] is compared with 
the expected histogram for full-flux preservation during cooling, where the vortex count per $x$-bin scales as $B_{\mathrm{FC}} \times (\text{Bin size})$ (horizontal green line). 
We confirm good agreement in the central region, whereas occurrence at the edge is suppressed.

\begin{figure*}[tbp]
  \centering
  \includegraphics[width=\textwidth]{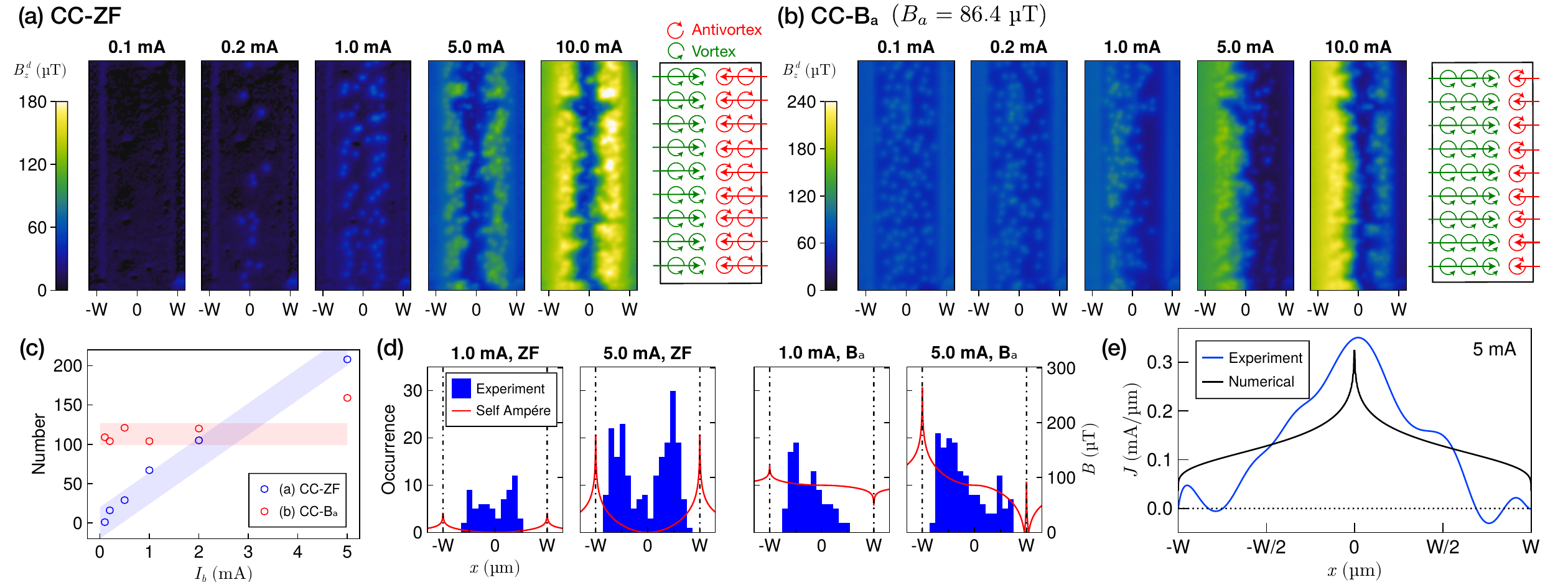}
  \caption{%
    (a) QDM maps after CC-ZF at $T = 71.4~\si{K}$ for different currents.
    (b) QDM maps after CC-$\mathrm{B_a}$ ($B_a=\SI{86.4}{\micro\tesla}$) at $T = 71.4~\si{K}$ for different currents.
    (c) Vortex count versus current for the datasets in (a) and (b).
    (d) Histograms of the vortex $x$-positions at  1.0~\si{mA} and 5.0~\si{mA} for (a) and (b). The solid red curves are the distributions of the magnitude of the Amp\`ere self-field, with magnetic flux density rescaled to the vortex occurrence.
    (e) Current distribution inferred from the CC-ZF (5.0~\si{mA}) data (blue), compared with the numerical self-consistent solution (black) [\cref{eq:self-consistent}].
  }
  \label{fig:CC}
\end{figure*}

We define two regions of interest (ROIs) inside the strip: ROI-I (full width, $|x|\leq W$) and ROI-II (central band, $|x| \leq 10~\si{\micro m}$), drawn in \cref{fig:FC}(a).
\Cref{fig:FC}(b) plots the vortex-carried flux density $N^{\mathrm{ROI}}\phi_0/S^{\mathrm{ROI}}$ in each ROI as a function of $B_{\mathrm{FC}}$, where $N^{\mathrm{ROI}}$ is the number of vortices in the ROI, $\phi_0$ is the flux quantum, and $S^{\mathrm{ROI}}$ is the ROI area.
A clear proportionality is observed with scale factor $\xi \sim 0.74<1$ in ROI-I. In contrast, ROI-II yields $\xi'\sim 1.03 \approx 1$, indicating flux conservation in the interior and depletion near the ends. 
This trend, where the vortex nucleation is suppressed within the edge width of $\delta\sim W/4 \sim\SI{4}{\micro m}$ agrees with previous reports~\cite{Embon2017,Veshchunov2016} and with the geometrical barrier's behavior~\cite{Zeldov1994Geometrical}.

\textit{Current-biased cooling}---Using the same device, we next examine the magnetic-field distributions produced by CC at zero-applied magnetic field (CC-ZF). 
The sample is imaged after cooling to \SI{71.4}{K} while maintaining $I=0.1\text{--}5.0~\si{\milli\ampere}$. \Cref{fig:CC}(a) shows the QDM maps: no vortices appear at $0.1~\si{\milli\ampere}$, whereas vortex nucleation is observed at $0.2~\si{\milli\ampere}$ and above. 
Up to $5.0~\si{\milli\ampere}$, individual vortices remain well separated, enabling position identification and counting~\cite{SI}. 
As shown in \cref{fig:CC}(c), the total vortex count increases linearly with current. 
Notably, as the current increases, \emph{vortices increasingly avoid the strip center} in \cref{fig:CC}(a). 

We interpret this center-avoiding distribution as reflecting the magnetic-field profile in the vortex-flow state driven by the self-field.
With our method, the sign of $B^{\,d}$ cannot be determined near zero-applied field, so we infer polarity from geometry.
When the current is driven upward [see \cref{fig:schematic}(b)], the self-field [\cref{eq:biot-savart}] gives $B>0$ at the left edge and $B<0$ at the right.
At temperatures where vortices can flow, the Lorentz force on vortices of either sign points toward the strip interior.
If vortices nucleate under this situation, as sketched to the right in \cref{fig:CC}(a), positive-polarity vortices flow rightward, with the left edge acting as a source and the right as a sink; antivortices flow leftward with the roles reversed.
Near the center, the two species meet and annihilate.
In this picture, the center-avoiding vortex distribution reflects the self-field symmetry: vortex nucleation is suppressed near the center by symmetry, where the total magnetic field vanishes. 
As discussed later, this self-field-driven flux flow plays a key role in shaping the final current distribution.

Cooling in a finite applied field $B_a$ (CC-$B_a$) yields a different pattern. 
With $B_a=\SI{86.4}{\micro\tesla}$, we cool to \SI{71.4}{K} while maintaining $I=0.1\text{--}5.0~\si{\milli\ampere}$ [\cref{fig:CC}(b)]. 
At $0.1~\si{\milli\ampere}$, FC-induced nucleation dominates and vortices are nearly uniform across the strip [polarity out of the page; colored green in \cref{fig:CC}(b)]. 
For $I\ge 0.2~\si{\milli\ampere}$, the total count remains close to that at $0.1~\si{\milli\ampere}$, but the distribution shifts toward the left, i.e., \emph{opposite to the Lorentz-force direction} indicated by green arrows in the right schematic of \cref{fig:CC}(b). 
At higher currents ($5.0$ and $10.0~\si{\milli\ampere}$), a depletion valley develops to the right of center.

\begin{figure}[htbp]
\centering
\includegraphics[width=0.5\textwidth]{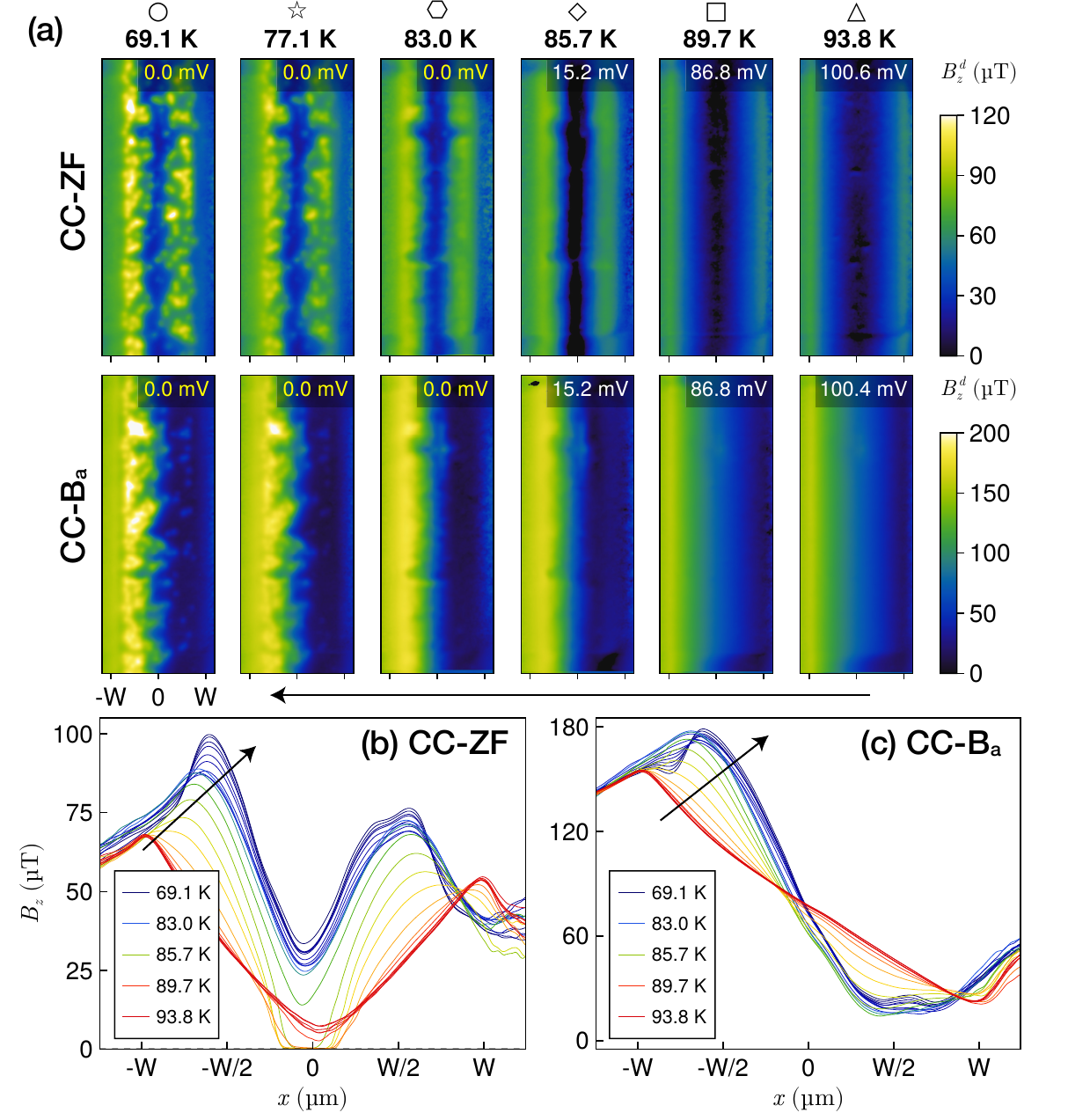}
\caption{%
(a) QDM maps obtained at different temperatures during cooling from \SI{93.8}{K} to \SI{69.1}{K} under CC-ZF (top) and CC-$B_a$ ($86.4~\si{\micro\tesla}$, bottom). The arrow at the bottom indicates measurement order. The vortex configuration is unchanged after setting the current to $0~\si{\milli\ampere}$ following the cooldown~\cite{SI}.
(b-c) Cross-sectional profiles averaged along the $y$ direction for each map. Traces are taken at every $1~\si{\kelvin}$, and representative traces are shown in the legend.
Panels (b) and (c) correspond to CC-ZF and CC-$B_a$, respectively.
The arrows indicate the measurement order.
}
\label{fig:CC-Tdep}
\end{figure}

These observations can again be qualitatively understood as a consequence of the self-field. 
For currents where $B_a$ dominates the self-field, \cref{eq:biot-savart} predicts a left--right asymmetry in the magnetic-flux-density profile: because $B_a$ adds to the self-field, the profile decreases monotonically toward the right, so the left side has the larger magnitude (same polarity) and nucleates more vortices. 
The observed leftward-biased vortices thus provide an evidence for the role of the self-field.
When the self-field exceeds $-B_a$, the total field reverses sign. 
In contrast to the CC-ZF case, where the sign reversal occurs at the center, the presence of a finite $B_a$ shifts the location where the total field vanishes to the right [compare \cref{fig:CC}(a,b) at $I=5.0$ and $10.0~\si{\milli\ampere}$].

\textit{Current distribution}---While the self-field sets the overall vortex distribution, the data deviate qualitatively from the self-field expected for a uniform current [\cref{eq:self-ampere}]. \Cref{fig:CC}(d) shows histograms of the vortex $x$-positions for CC-ZF and CC-$\mathrm{B_a}$ at $1.0$ and $5.0~\si{\milli\ampere}$. 
The solid red curves show the absolute value of the self-field expected for a uniform current [\cref{eq:self-ampere}]. 
Although overall shapes are similar, the histogram peaks lie significantly inward of the self-field peaks, which are at the edges, indicating a nonuniform, center-biased current. 
We reconstruct the current distribution that reproduces the QDM map at $5.0~\si{\milli\ampere}$ in CC-ZF (see \cref{sec:numerical,sec:current_estimate}).
The resulting distribution, shown as the blue curve in \cref{fig:CC}(e), exhibits a characteristic profile with a sharp central peak.

Furthermore, we acquire the QDM maps at different temperatures during cooldown, with the current fixed at \SI{5.0}{\milli\ampere}, as shown in \cref{fig:CC-Tdep}(a) [CC-ZF (top) and CC-$\mathrm{B_a}$ (bottom)]. The simultaneously measured time-averaged voltages [see \cref{fig:schematic}(b)] are annotated at the top of each panel. In both cases, as $T$ decreases, the flux-density peaks shift inward and eventually halt around the temperature at which the voltage drops to zero, namely between $T=85.7~\si{\kelvin}$ and $T=83.0~\si{\kelvin}$.

\begin{figure}[tbp]
\centering
\includegraphics[width=0.5\textwidth]{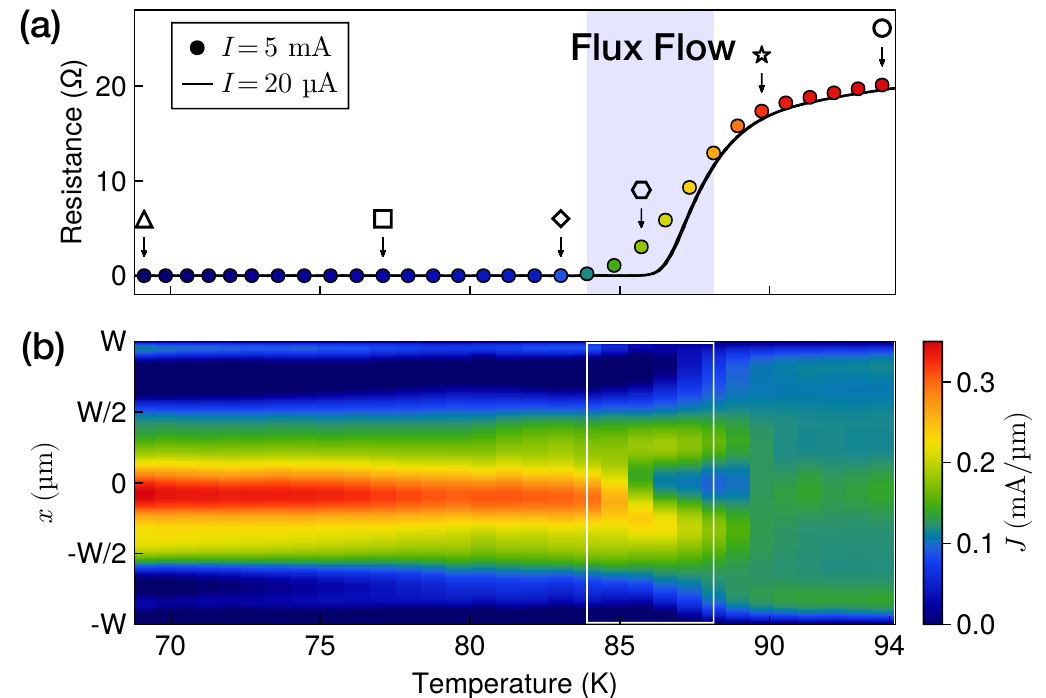}
\caption{%
(a) Temperature dependence of the resistance at $20~\si{\micro\ampere}$ and $5~\si{\milli\ampere}$. Markers correspond to the panels in \cref{fig:CC-Tdep}(a), and the colors of the filled markers match those of the traces in \cref{fig:CC-Tdep}(b).
(b) Current distributions as a function of $T$ reconstructed from the CC-ZF results in \cref{fig:CC-Tdep}(a). The white box indicates the blue band in (a).
}
\label{fig:CC-Tdep-RT-I}
\end{figure}

The measured voltages divided by $I=\SI{5.0}{\milli\ampere}$ are plotted in \cref{fig:CC-Tdep-RT-I}(a), overlaid with the temperature–resistance curve at \SI{20}{\micro\ampere} excitation current~\footnote{These voltage data were acquired in zero-applied field; no significant differences were found with an applied field.}. 
A clear deviation between the \SI{20}{\micro\ampere} and \SI{5.0}{\milli\ampere} traces appears as $T$ decreases through the range 85.7--83.0~$\si{\kelvin}$, as indicated by a blue band. Notably, while the \SI{20}{\micro\ampere} trace reaches zero resistance around $86~\si{\kelvin}$, a finite voltage remains at \SI{5.0}{\milli\ampere}. This indicates a resistive contribution from vortex flow, supporting the self-field-driven scenario in \cref{fig:CC}(a).

\Cref{fig:CC-Tdep}(b,c) show the $y$-averaged QDM profiles obtained at different temperatures. One sees that, in both CC-ZF and CC-$\mathrm{B_a}$, the peak shift becomes most pronounced within this flux-flow regime highlighted by the blue band in \cref{fig:CC-Tdep-RT-I}(a). \Cref{fig:CC-Tdep-RT-I}(b) shows the current distributions reconstructed from the CC-ZF profiles in \cref{fig:CC-Tdep}(b) as a function of $T$. The current evolves from uniform to center-biased as $T$ falls, with the strongest changes occurring within the flux-flow regime.

\textit{Theoretical treatment}---We observe current concentrating near the strip center in CC-ZF. What mechanism causes the current to concentrate at the strip center during vortex nucleation and freezing? Thus far, we have implicitly assumed that the vortex distribution is set solely by the self-field, which maintains the local magnetic-flux density after cooldown, while the current density remains uniform. This assumption, however, neglects the feedback of vortex motion on the current density. As noted in the introduction, $J(x)$ and $B(x)$ are related nonlocally and self-consistently via \cref{eq:biot-savart}, so the influence of vortex motion on $J(x)$ is not \emph{a priori} obvious.

Here, based on the flux-flow~\cite{Bardeen1965,Kim1965,Brandt1995} and flux-creep models~\cite{Kim1962,Blatter1994,Mikitik2013,Brandt1996,Gaevski1999,Yeshurun1996}, we quantitatively examine the dynamics up to vortex freeze-out and the resulting flux density and current profiles, focusing on CC-ZF. In a quasi-1D system, the out-of-plane magnetic field $B$ and the $y$-component of the electric field $E$ satisfy
\begin{equation}
  \pdv{B}{t} = -\pdv{E}{x}\,.
  \label{eq:flux-continuity}
\end{equation}
In the normal state with uniform Ohmic sheet resistivity $\rho_n$, the electric field is $E=\rho_n J_0$ for a uniform sheet current $J_0$.
In the flux-flow state, following \citet{Kim1965}, the electric field arises from vortex motion in the $x$-direction with velocity $v$.
Balancing the Lorentz force on a vortex, $\vec J \times \vec \phi_0$, against viscous drag, $-\eta v$, yields
\begin{equation}
  E = v B = \frac{B \phi_0}{\eta}\, J \equiv \rho_{\mathrm{ff}} J,
  \label{eq:rho-ff}
\end{equation}
where $\rho_{\mathrm{ff}}$ is the flux-flow sheet resistivity.
Given upper critical field $B_{c2}(T)$, when $B < B_{c2}(T)$, $\rho_{\mathrm{ff}} \approx \rho_n\, B / B_{c2}(T)$, whereas for $B > B_{c2}(T)$ the response crosses over to the normal-state value $\rho_n$~\cite{Brandt1995,Kim1962,Kunchur1993}.

We now consider when \cref{eq:flux-continuity} reaches a steady state, $\partial B/\partial t=0$, which implies a position-independent electric field, $E(x)\equiv \mathrm{const}$. In the normal state, $\rho(x)\equiv\rho_n$, and therefore $J(x)\equiv \mathrm{const}$. Once the strip enters the flux-flow state, the self-consistency condition
\begin{equation}
  B[J](x)\,J(x)=\mathrm{const}
  \label{eq:self-consistent}
\end{equation}
follows from $E=\rho_{\mathrm{ff}}J\propto BJ$. As $\rho$ evolves from $\rho_n$ to $\rho_{\mathrm{ff}}$, the response crosses over from normal conduction to flux flow. \Cref{eq:self-consistent} then implies that, in self-field-driven flux flow, the current concentrates near the strip center, where $B\approx 0$ for CC-ZF. On further cooling the system crosses into flux creep, where $v \propto [J/J_c]^n$ with $J_c$ a characteristic  current and $n=U_c/(k_BT)$ in terms of the pinning-energy scale $U_c$ and Boltzmann's constant $k_B$ (see \cref{sec:v-j-creep}), yielding the power-law generalization
$B[J](x)\,\bigl[J(x)/J_c\bigr]^n=\mathrm{const}.$
In this regime, flux is eventually pinned stochastically during cooldown, so the field pattern effectively freezes, producing a glassy state.

With this understanding, we re-examine the CC-ZF data. 
In \cref{fig:CC}(e), the black curve shows the numerical solution of \cref{eq:self-consistent} (for the computational procedure, see \cref{sec:numerical}), which reproduces the sharp central cusp in the experimental profile (blue). 
\Cref{fig:CC-Tdep-RT-I}(b) shows that, as $T$ decreases, this cusp develops at the onset of flux-flow resistance and changes little once zero resistance is reached.
Values $1 \le n \le 2$, corresponding to flux flow and flux creep near the critical point, describe the data well, whereas larger $n>5$ do not \cite{SI}. 
These observations suggest that the vortex configuration freezes early in the flux-creep regime during cooldown.

Closer inspection of \cref{fig:CC}(e) reveals that the experimental profile is broader at the center and steeper near $x=\pm W/2$ than in the model. Two effects can shift the flux peaks inward. 
(i) The analysis above does not distinguish the entry barrier at the edge from the hopping barrier between bulk pinning sites; the entry barrier can be larger. A temperature window may therefore exist in which the edge nucleation rate and the bulk hopping rate decouple, depleting edge vortices. 
(ii) As seen in \cref{fig:FC}, a region of width $\delta \sim W/4$ near the edges disfavors vortex occupancy due to the edge barrier~\cite{Zeldov1994Geometrical,Bean1964surface}, further suppressing their presence. 
Both effects shorten the mean residence time of vortices near the edges. Viewing the measured flux density as the product of a vortex's stray field and its residence time~\cite{Embon2017}, the reduced residence time lowers the edge signal, accounting for the discrepancy between the experiment and the numerical solution in \cref{fig:CC}(e).

\textit{Conclusion}---
We examined the crossover from the normal state, through flux flow, to a frozen state under current-biased cooling (CC) in a hard type-II superconducting strip.
Even at zero-applied field, CC generates vortices whose number grows linearly with current. 
Under a small applied field, CC yields the vortex distribution shifted opposite to the Lorentz-force direction.
These observations are explained by the Amp\`ere self-field.
We also observe current concentrating at the strip center during cooldown, accompanied by the onset of flux-flow resistance near the transition.
These results are consistent with steady-state flux-flow solutions and their temperature-driven crossover.

Even currents small enough not to suppress the superconducting transition generate vortices that retain history.
This motivates a reassessment of vortex capture in current hysteresis of superconductors.
In dc transport of similar thin films, large dc currents produce flux-flow resistance driven by self-field; upon reducing the current, the induced vortices can be trapped and frozen across the strip, reflecting the field distribution of flux-flow steady state.
Such asymmetric vortex distributions may lead to unusual nonreciprocal transport.

More broadly, our measurements directly probe a driven particle system that undergoes a glassy transition upon cooling, offering new experimental guidance for nonequilibrium statistical physics via vortex matter. Comparisons with pulsed or ac drives~\cite{Okuma2011} should enable cross-disciplinary tests of dynamic glass transitions across driven systems.

\textit{Acknowledgement}---We thank Yusuke Kato, Tsuyoshi Tamegai, Kazuaki Takasan, and Daiki Nishiguchi for fruitful discussions. 
This work is partially supported by JST, CREST Grant Number JPMJCR23I2, Japan; JSPS Grants-in-Aid for Scientific Research (Nos. JP23K25800, JP22K03524, JP22KJ1059, JP22J21412, JP24K21194, JP25K00934, JP24KJ0657, and JP25H01248); 
MEXT Quantum Leap Flagship Program (MEXT Q-LEAP) Grant Number JPMXS0118067395;
Seiko Instruments Advanced Technology Foundation Research Grant; 
``Advanced Research Infrastructure for Materials and Nanotechnology in Japan (ARIM)'' of the Ministry of Education, Culture, Sports, Science and Technology (MEXT), Proposal Number JPMXP1222UT1131 and JPMXP1225NM0125; 
Daikin Industries, Ltd; 
the Cooperative Research Project of RIEC, Tohoku University; 
New Challenge Research hosted by JSR Corporation via JSR-UTokyo Collaboration Hub, CURIE.
S.N. is supported by the Forefront Physics and Mathematics Program to Drive Transformation (FoPM), WINGS Program, and JSR Fellowship, the University of Tokyo.
\appendix
\setcounter{secnumdepth}{2}

\section{Methods for numerical calculation}
\label{sec:numerical}
The numerical solutions for the self-consistent equation
\begin{equation}
  B[J](x) J(x)=\text{Const.}
  \tag{\cref{eq:self-consistent} in the main text}
\end{equation}
can be obtained as follows.
We expand the current distribution $J(x)$ in weighted Chebyshev polynomials of the first kind,
\begin{equation}
  J(x)=\frac{1}{\sqrt{1-\tilde x^{2}}}\sum_{n\ge 0} a_n T_n(\tilde x),\qquad \tilde x = x/W .
  \label{eq:current}
\end{equation}
For the finite Hilbert transform
\begin{equation}
  \mathcal H[f](x)=\frac{1}{\pi} \pdvd \int_{-1}^{1}\frac{f(u)}{u-x} du\ \text{for}\ |x| \leq 1,
\end{equation}
one has~\cite{TableofIntegrals,Huang2022,Steinberger2019}
\begin{gather}
  \mathcal H\left(\frac{T_n(x)}{\sqrt{1-x^{2}}}\right)=U_{n-1}(x),
\end{gather}
where $U_n$ are Chebyshev polynomials of the second kind. Hence
\begin{equation}
  B(x)=\frac{\mu_0}{2}\sum_{n\ge 0} a_n U_{n-1}(\tilde x),
\end{equation}
which expresses $B$ directly in terms of the coefficients $a_n$. We set $U_{-1}(x)\equiv 0$.
The inverse transform $\mathcal H^{-1}$ is given by the inversion formula~\cite{Tricomi1951},
\begin{align}
  \mathcal H^{-1}[f](x)=&\frac{1}{\pi} \pdvd \int_{-1}^{1} \frac{f(u)}{u-x}\sqrt{\frac{1-u^{2}}{1-x^{2}}} du\notag\\
  &+\frac{C}{\sqrt{1-x^2}}\ \text{for}\ |x| \leq 1.
\end{align}
Here, the finite Hilbert transform $\mathcal H$ has a one-dimensional nullspace spanned by $(1-x^{2})^{-1/2}$, which produces the additive ambiguity $C/\sqrt{1-x^{2}}$. The constant $C$ can be fixed by an external constraint, e.g. by imposing the total current~\cite{Norris1970,Brandt1993,Zeldov1994Magnetization}. 
With the convention $\displaystyle \mathcal H\left(\frac{T_n(x)}{\sqrt{1-x^{2}}}\right)=U_{n-1}(x)$, the inverse satisfies, for $n\ge1$,
\begin{equation}
  \mathcal H^{-1}\left(T_n(x)\right)
  = -\sqrt{1-x^{2}}U_{n-1}(x),
\end{equation}
while for $n=0$ one has $\mathcal H^{-1}[T_0]=C/\sqrt{1-x^{2}}$ due to the nullspace term. Chebyshev polynomials obey
\begin{align}
  T_n(\tilde x) &= \cos(n\theta),\\
  U_n(\tilde x) &= \frac{\sin\big((n+1)\theta\big)}{\sin\theta},\qquad \theta=\arccos(\tilde x) .
\end{align}
Sampling at
\begin{equation}
  x_k = W\cos\left(\frac{2k+1}{2N}\pi\right)=W\cos\theta_k,\quad k=0,1,\dots,N-1 ,
\end{equation}
the relation \cref{eq:current} becomes
\begin{align}
  J(x_k)\sqrt{1-\tilde x_k^{2}}
    &= \sum_{n\ge 0} a_n T_n(\tilde x_k)\notag\\
    &= a_0 + \sum_{n\ge 1} a_n \cos\left(n\frac{2k+1}{2N}\pi\right) \notag\\
    &= \sqrt{N} \mathcal D^{-1}[a_n](k),
\end{align}
that is, a discrete cosine transform (DCT) of type III, $\mathcal D^{-1}$. 
The coefficients $a_n$ are then obtained with the type-II DCT, $\mathcal D$,
\begin{align}
  a_n &= \mathcal D\big[J(x_k)\sin\theta_k\big](n) \Lambda_n, \label{eq:cheby_expansion}\\
  \Lambda_n &=
  \begin{cases}
    \sqrt{1/N}, & n=0,\\
    \sqrt{2/N}, & \text{otherwise}.
  \end{cases}\notag
\end{align}
Useful derivatives are
\begin{align}
  \dv{T_n(x)}{x} &= n U_{n-1}(x), \notag\\
  \dv{U_n(x)}{x} &= \frac{(n+1)T_{n+1}(x)-x U_n(x)}{x^{2}-1}.
\end{align}

With these ingredients, one supplies an initial guess for $a_n$ from \cref{eq:cheby_expansion} using a suitable $J(x)$ profile, for example the constant $J$ corresponding to the right-hand side of \cref{eq:self-consistent}, then solves \cref{eq:self-consistent} as a nonlinear system. Because analytical derivatives are available, evaluating the Jacobian is straightforward, so we employ a trust-region method for the solver.

Moreover, in the present case for zero-applied field, symmetry requires an even current distribution, $J(x)=J(-x)$, while the self-field in \cref{eq:biot-savart} is then an odd function. Because the sign reverses at $x=0$, one must have $B(x=0)=0$. If this is imposed self-consistently, $J$ diverges at $x=0$. On the other hand, as discussed in the main text, in a self-field-driven flux-flow state vortices and antivortices move in opposite directions and collide in some region. If this collision region is, on average, at $x=0$, creation and annihilation of flux there allow a local breakdown of the continuity equation \cref{eq:flux-continuity}.

\section{Estimating the Current from Magnetic-Field Measurements}
\label{sec:current_estimate}
Current reconstruction from field data is often performed with Fourier-transform methods~\cite{Roth1989,Zuber2018,Broadway2020,Johansen1996,Gaevski1999}. 
Near zero-applied field, however, our method cannot determine the sign, and nonuniformity along the strip length makes longitudinal averaging a biased estimator. 
Taking these points into account and for flexibility in penalizing errors in the magnetic field near the strip center and handling absolute-value operations, we compute the actual magnetic-field profile from the current as Chebyshev series in \cref{eq:current}, and then estimate the current via least-squares optimization.

\Cref{eq:biot-savart} gives the field just above the superconductor, whereas NV centers sense the field at a finite stand-off distance. For a separation $d$, the out-of-plane component $B^{\,d}(x)$ at $(x,d)$ follows from the Biot-Savart law as
\begin{equation}
{
B^{\,d}(x)=\frac{\mu_0}{2\pi}\int_{-W}^{W}\frac{t-x}{(t-x)^2+d^2}J(t)dt.
}
\end{equation}

Here we consider the Cauchy transform, which extends the Hilbert transform to the complex plane,
\begin{equation}
\mathcal C[J](z)
=\frac{1}{\pi}\int_{-1}^{1}\frac{J(s)}{s-z}ds,\qquad z\in\mathbb C\setminus[-1,1].
\end{equation}
Setting $z = x + id$ gives
\begin{equation}
\frac{1}{s-(x+id)}=\frac{s-x}{(s-x)^2+d^2}+i\frac{d}{(s-x)^2+d^2}.
\end{equation}
Therefore,
\begin{equation}
B^{\,d}(x) = \frac{\mu_0}{2}\Re\mathcal C[J](x + id).
\end{equation}
At this point we can again apply the Chebyshev-polynomial expansion method from the previous section.
\begin{equation}
\eta = \operatorname{arccosh} z = \log\bigl(z + \sqrt{z^2 - 1}\bigr).
\end{equation}
Then, defining
\begin{align}
\beta(z) &= \sqrt{z^2 - 1} = \sinh \eta, \notag\\
\gamma(z) &= z - \sqrt{z^2 - 1} = \bigl(z + \sqrt{z^2 - 1}\bigr)^{-1} = e^{-\eta},
\end{align}
we obtain
\begin{equation}
\mathcal C \qty[\frac{T_n(s)}{\sqrt{1-s^2}}] (z)
 = -\frac{\gamma^n(z)}{\beta(z)} = -\frac{e^{-n\eta}}{\sinh \eta}.
\end{equation}
Hence, for $z=x+i d$,
\begin{equation}
B^{\,d}(x) = -\frac{\mu_0}{2}\sum_{n\geq 0} a_n\frac{e^{-n\eta}}{\sinh \eta}.
\end{equation}
Using the coefficient sequence $\{a_j\}$, the magnetic field $B^{\, d}_z(x_k)$ at $x_k$ becomes
\begin{align}
B^{\,d}(x_k)
&=\frac{\mu_0}{2}\Re\mathcal C[J](x_k + id)\notag\\
&=-\frac{\mu_0}{2}\sum_{j\geq 0} a_j\frac{e^{-j\eta_k}}{\sinh \eta_k}\notag\\
&= K_{kj} a_j,
\end{align}
where we set
\begin{equation}
K_{kj} = -\frac{\mu_0}{2}\frac{e^{-j\eta_k}}{\sinh \eta_k},\ \eta_k = x_k + id.
\end{equation}
Such a kernel-based discretization is similar to that of \citet{Brandt1996}; it is accurate for a given Chebyshev-series truncation order. Because the result is given by multiplication with a precomputed kernel, it is computationally advantageous. Moreover, the finite stand-off distance suppresses the edge divergences, improving numerical stability. We obtain the magnetic field from this kernel-based transform and fit the field shape by least squares.

\section{Derivation of $v$--$J$ Power Law}
\label{sec:v-j-creep}
Here, based on the flux-creep model~
\cite{Kim1962,Yeshurun1996,Blatter1994,Mikitik2013,Brandt1996,Gaevski1999}, we analyze the transient motion until vortices freeze. Assuming that hopping occurs when the Lorentz force probabilistically overcomes the pinning energy $U$, the relation between the vortex velocity $v$ and the current density $J$ obeys an Arrhenius form,
\begin{equation}
  v(j,T)=v_c\exp\bigl[-U/k_{\mathrm B}T\bigr].
  \label{eq:E-j}
\end{equation}
Here $U$ is the activation barrier. In the flux-creep model, $U$ depends on the local current density, $U=U(J)$, whose form is system dependent. We assume a logarithmic relation
\begin{equation}
  U(j)=U_{c}\log(j_{c}/j),
  \label{eq:U-j0}
\end{equation}
which, in the parameter range of our measurements, is known to approximate well the more rigorous power-law result with $\mu=1/7$ for single-vortex collective pinning~\cite{Blatter1994,Yeshurun1996}. In this case
\begin{equation}
  v(J,T)=v_c (J/J_c)^{n}, \qquad n=\frac{U_c}{k_B T}.
\end{equation}

%% file: letter_SI_nightly.tex
\pagebreak
\widetext
\setcounter{section}{0}
\renewcommand{\thesection}{S-\Roman{section}}
\setcounter{equation}{0}
\setcounter{figure}{0}
\setcounter{table}{0}

\makeatletter
\renewcommand{\theequation}{S\arabic{equation}}
\renewcommand{\thefigure}{S\arabic{figure}}
\renewcommand{\bibnumfmt}[1]{[S#1]}
\renewcommand{\citenumfont}[1]{S#1}

\section{Details of the Theoretical Model}
\subsection{Quasi-1D geometry for flux flow and flux creep}
\begin{figure}[htbp]
  \begin{center}
    \includegraphics[width =0.5\textwidth]{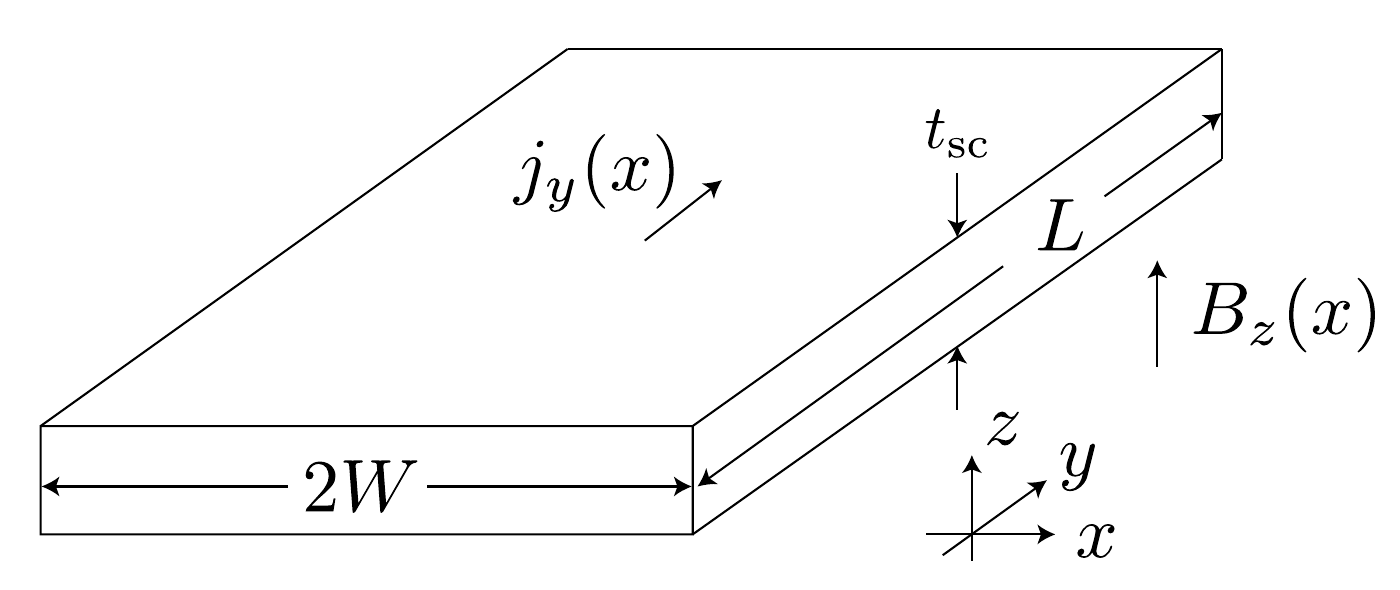}
    \caption{
      Diagram of a thin-film strip.
    }
    \label{fig:diagram}
  \end{center}
\end{figure}

In the main text, we interpreted both the frozen vortex distribution and the transient flux-flow state as a crossover between the flux-flow resistivity model~\cite{si_Blatter1994,si_Bardeen1965,si_Kim1965,si_Brandt1995} and the flux-creep model~\cite{si_Kim1962,si_Blatter1994,si_Mikitik2013,si_Brandt1996,si_Gaevski1999,si_Yeshurun1996}. On this basis we used
\begin{equation}
  B[J](x) J(x)=\text{const} \tag{Eq.~(5) in the main text}
\end{equation}
to interpret the current distribution during the flowing state. 
To clarify the assumptions underlying this model, we first derive the quasi-1D sheet relation from the 3D geometry, following \citet{si_Brandt1993}. We use the same coordinate system as in the main text, see \cref{fig:diagram}.

The Biot-Savart law for the volumetric current density $\vec{j}$ reads
\begin{equation}
  \vec B (\vec r) = \frac{\mu_0}{4\pi} \int\frac{\vec j (\vec r) \times (\vec r - \vec r')}{|\vec r - \vec r'|^3} d\vec r',
  \label{eq:biot-savart-full}
\end{equation}
where $\mu_0$ is the vacuum permeability. For a sufficiently long strip ($L\gg 2W$) we have translational symmetry along $y$, hence $\vec{j}(\vec r') = j_y (x', z') \vec e_y$. Performing the $y'$ integral reduces \cref{eq:biot-savart-full} to a 2D geometry,
\begin{align}
  \vec B (x, z) 
  &= \frac{\mu_0}{4\pi} \int dx' \int dz' 
  [(x'-x)\vec{e}_z + z\vec e_x] j_y (x', z')
  \int_{y' \in \mathbb R} \frac{dy'}{\qty[(x-x')^2 + (y')^2 + (z-z')^2]^{3/2}} \notag\\
  &= \frac{\mu_0}{2\pi} \int dx' \int dz' \frac{[(x'-x)\vec{e}_z + z\vec e_x]}{(x-x')^2 + (z-z')^2} j_y(z')
\end{align}
We distinguish the volumetric current density $\vec j$ from the sheet current density along $y$ at position $x$, denoted $J(x)$, defined by
\begin{equation}
  J(x) = \int_{-t_\mathrm{sc}/2} ^{t_\mathrm{sc}/2} j_y(x, z)dz.
\end{equation}
In the thin-film limit we replace the thickness profile by a sheet,
\begin{equation}
  j_y(x, z) = J(x)\delta(z)dz,
\end{equation}
which yields
\begin{align}
  \vec B (x, z) 
  &= \frac{\mu_0}{2\pi} \int_{-W}^W \frac{(x'-x)\vec{e}_z + z \vec e_x}{(x-x')^2 + z^2}  J(x') dx'.
\end{align}
At the surface of the superconductor $z \to 0^+$, we obtain
\begin{align}
  B_z (x) 
  = \frac{\mu_0}{2\pi} \mathrm{p.v.} \int_{-W}^W  \frac{J(x')}{x'-x} dx' \quad (\text{Eq.~(1) in the main text}), \quad B_x (x) = \frac{\mu_0 J(x)}{2}. 
\end{align}
These relations hold with accuracy $O(t_{\mathrm{sc}}/W)$~\cite{si_Brandt1993} as long as translational symmetry along $y$ is valid.

We now consider the flux-flow constitutive relation. In this regime, the in-plane field $B_x$ is odd in $z$, so its thickness average inside the film vanishes by symmetry.
Here vortices are also taken to be perpendicular to the surface. 
The Lorentz force per unit length is $\phi_0 j_y$; integrating across the thickness gives the total force $\phi_0 J(x)$. For the viscous drag, integrating across the thickness yields the drag force $\eta v(x)$. Balancing the two yields $v(x)=\frac{\phi_0}{\eta}J(x).$ The electric field is parallel to the current, and in the quasi-1D reduction it has only a $y$ component inside the superconductor, yielding
\begin{equation}
  E_y(x,z) = v(x) B_z(x, 0) = \frac{B_z \phi_0}{\eta} J \equiv \rho_{\mathrm{ff}} J. \tag{Eq.~(4) in the main text}
\end{equation}
which is the sheet version of the flux-flow law. 
Here $\rho_{\mathrm{ff}}=B_z \phi_0/\eta$ is the \emph{sheet} flux-flow resistivity.
In the zero applied field case of interest, the $B_z$ that enters $\rho_{\mathrm{ff}}$ arises solely from the self-field, so the problem reduces to the self-consistent equation [Eq.~(5) in the main text]. When a finite applied field $B_a$ is present, the flux-flow resistivity is
\[
\rho_{\mathrm{ff}} = (B_z + B_a)\,\frac{\phi_0}{\eta}.
\]
Note that the unit of $\eta$ and $\rho_\mathrm{ff}$ here differ from \citet{Kim1962}, which formulates volumetric relation.

\subsection{Comparison between the $v$--$J$ and $E$--$J$ models}
For flux creep there are two formulations: one assumes that the vortex velocity $v$ obeys an Arrhenius law, which we adopt in the main text following \citet{Kim1962} and \citet{Anderson1962}, and the other assumes that the electric field $E$ obeys an Arrhenius law, following \citet{Brandt1996} and \citet{Mikitik2013}.
To our knowledge, only \citet{Shantsev2000comparison} discusses the difference between these approaches at the level of the magnetic-field distribution.
Hereafter, we call the former the $v$--$J$ model and the latter the $E$--$J$ model, and we compare them in light of our experimental results.

Here, following \citet{Brandt1996}, we derive the time evolution of the magnetic flux and the current distribution via the $E$--$J$ formulation.
Among Maxwell's equations, Amp\`ere's law for the vector potential oriented along $y$ in the $x\text{\hyphen}z$ plane gives
\begin{equation}
  \mu_0 j(\vec r) = -\nabla^2 A(\vec r),
\end{equation}
and Faraday's law gives
\begin{equation}
 \nabla \times \vec E = -\pdv{t} \vec B = -\nabla \times \pdv{t}\vec{A}.
\end{equation}
Taking the gauge $x B_a$ with regard to the externally applied field $B_a$, we obtain
\begin{equation}
\vec E = -\pdv{t} \vec A .
\end{equation}
The time evolution of the crrent distribution can then be written in terms of the electric field as
\begin{equation}u
  \mu_0 \pdv{t} \vec j(\vec r,t) = \Delta \vec E(\vec r, t),
  \label{eq:time_evolv1}
\end{equation}
where, since $B_a$ is constant, the term $-x \dot B_a$ vanishes.
For the relation between the magnitude of the electric field $E$ and the current density $j$, we assume an Arrhenius form via the current dependence of the activation energy $U$,
\begin{equation}
  E(j,T)=E_c\exp\bigl[-U(j) /k_{\mathrm B}T\bigr],
  \label{eq:E-j}
\end{equation}
where $E_c$ denotes the characteristic electric field. 
As in the main text, a logarithmic pinning potential yields the power-law relation
\begin{equation}
  \mu_0 \pdv{t} j(r,t) = \Delta \left(E_c(j/j_c)^n\right), \quad  n = \frac{U_c}{k_{\mathrm B} T},
   \label{eq:time_evolv2}
\end{equation}
which is equivalent to a nonlinear diffusion equation.
If we impose boundary conditions that conserve the total current at both ends, the steady-state solution is characterized by $\pdv{j}{t}  = 0$, and, since the strip is mirror-symmetric, $E(x)$ is an even function.
Thus, \emph{independently of} $n$, the steady-state solution of \cref{eq:time_evolv2} is
\begin{equation}
  E(\vec r) = \lambda = \text{const.}
\end{equation}
When $j_c$ is spatially uniform, integrating over $z$ shows that this implies
\begin{equation}
  J(x) = \text{const.}
\end{equation}
From the above, in the steady state of flux creep, a uniform current $J(x)=\text{const.}$ is realized.
Since the normal-state current before the transition is also uniform, it is natural that this steady-state solution persists as the temperature changes, therefore the current is expected to remain uniform after cooldown.

\begin{figure}[tbp]
  \begin{center}
    \includegraphics[width =1.0\textwidth]{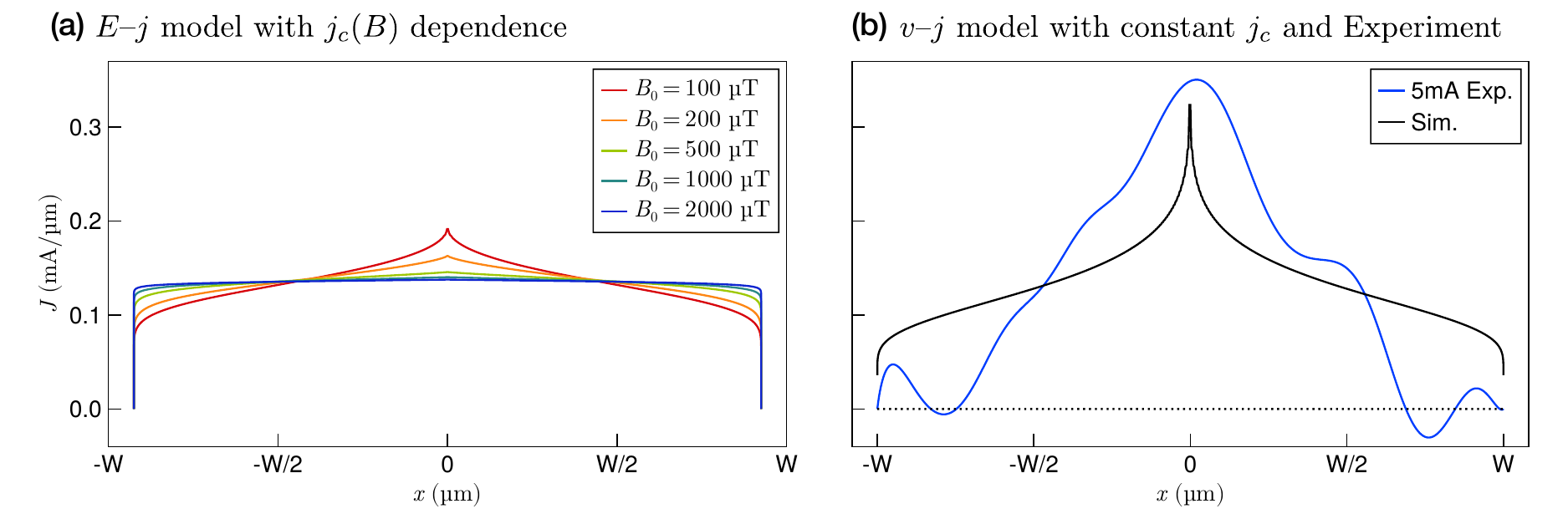}
    \caption{
      (a) Steady-state current distributions obtained in the $E$--$J$ model with the $J_c(B)$ dependence. (b) Reproduction of Fig.~5(b) from the main text.
    }
    \label{fig:E-j-kim}
  \end{center}
\end{figure}

This result therefore does not capture a crossover to the field-dependent behavior in the flux-flow state, and it does not reproduce our experimental data.
However, the conclusion of uniformity hinges on a local relation between $E$ and $j$, namely that $j(x')$ for $x'\neq x$ does not affect $E(x)$.
If, in any way, $E$ depends on $B$, this uniformity can break down.
As a possible cause of current crowding toward the center, we introduce a simple field dependence of the critical current, $J_c(B)=\dfrac{J_{c0}}{1+ B/B_0}$~\cite{si_Kim1962,si_Kim1964}, which yields the steady-state solution
\begin{equation}
  J(x)=\lambda J_{c0}\left(1 + \frac{1}{B_0} B_z[J](x)\right)^{-1},
  \label{eq:self-consistent2}
\end{equation}
rewritten self-consistently through the self-field.
Here we can choose $\lambda$ to set the total current.
In this self-consistent equation, the enhancement of the field near the edges suppresses $J_c(B)$ there, so $J(x)$ concentrates toward the center.
For fields much larger than $B_0$, \cref{eq:self-consistent2} reduces to the relation $B_z[J](x) J(x) = \text{const}$ in the main text.

The literature values of $B_0$ for cuprates are a few millitesla, smaller than those for conventional superconductors ($B_0 \gtrsim 50$~\si{\milli\tesla}~\cite{si_Kim1964})~\cite{si_Yasuoka1998,si_Mazaki1995}.
In our experiment, the self-field scale set by the current is $\mu_0 I/(2W) \sim \SI{169.8}{\micro\tesla}$, well below this millitesla range.
\Cref{fig:E-j-kim}(a) shows numerical solutions of the $E$--$J$ model's self-consistent equation [\cref{eq:self-consistent2}] for $B_0$ from $100$ to $2000$~\si{\micro\tesla}.
For comparison, \cref{fig:E-j-kim}(b) reproduces the main-text Fig.~3(e).
These calculations correspond to the case $n=1$ in the $v$--$J$ model.
Current distributions localized at the center, similar to the experiment, appear only for $B_0$ of several hundred~\si{\micro\tesla}.
Even for $B_0 = \SI{100}{\micro\tesla}$, the central peak is more suppressed than in the $v$--$J$ model, which better describes the experimental data.

\begin{figure}[tbp]
  \begin{center}
    \includegraphics[width =0.6\textwidth]{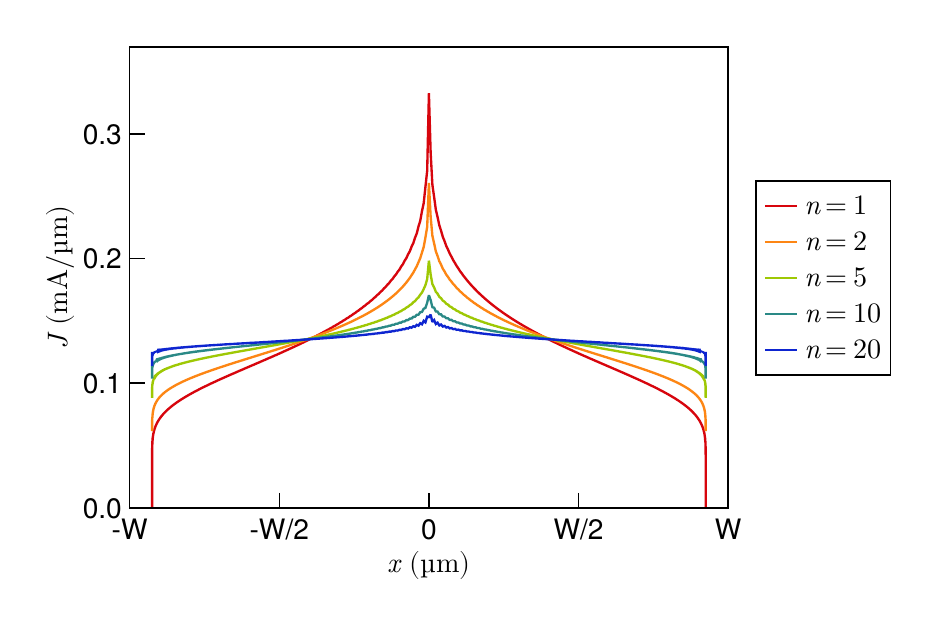}
    \caption{
      Dependence on $n$ in the $v$--$J$ model.
    }
    \label{fig:n-dep}
  \end{center}
\end{figure}

\subsection{On the crossover from flux flow to flux creep in the $v$--$J$ model}
In the $v$--$J$ model used in the main text,
\begin{equation}
  \dv{\vec B}{t} = \vec v \times \vec B,\quad v(J,T)=v_c \left(J/J_c\right)^{n}, \qquad n=\frac{U_c}{k_{\mathrm B} T},
\end{equation}
the power-law form incorporates, within one framework, the linear response $n=1$ characteristic of flux flow, the creep regime $n>1$, and, in the limit $n\to\infty$, the behavior of a frozen state akin to the Bean critical-state model~\cite{si_Brandt1996}.

Care is required in treating $U_c$.
For cuprates, reported experimental values of $U_c$ exhibit very large scatter~\cite{si_Hagen1989,si_Palstra1989,Gurevich1990,Gaevski1999}.
In \citet{Gaevski1999}\footnote{The main text of Ref.~\cite{si_Gaevski1999} lists $0.08$~meV, which contradicts the stated $\approx 50 k_{\mathrm B} T$; we regard $0.08$~eV as intended.} one finds
\begin{equation}
  U_c \approx 80~\si{meV} \approx 50 k_{\mathrm B} T \quad (\text{corresponding to } T \approx \SI{19}{K}),
\end{equation}
consistent with the median in \citet{Hagen1989}.

In cuprate superconductors at low temperatures, $T<30~\si{K}$, it has been observed that $U_c$ \emph{decreases} upon lowering the temperature; see, e.g., \citet{Campbell1990}.
At higher temperatures, one can expect, from relations between $U_c$ and fundamental superconducting parameters, that
\begin{equation}
  U_c(T)\simeq \frac{\phi_0^2}{8\pi \mu_0}\left(1-\frac{T}{T_c}\right),
\end{equation}
so $U_c$ vanishes linearly near $T_c$.
Alternatively, in the mean-free-path fluctuation scenario of \citet{Griessen1994}, $U_c\propto 1-(T/T_c)^4$.

Consider now a process in which the current is maintained while the temperature is lowered within the flux-creep regime.
Superconductivity first appears in the $n=1$ flux-flow state, then, according to \cref{eq:time_evolv2}, the system maintains a steady state as $n(T)$ increases rapidly below the transition, diffusion slows, and creep tends toward freezing.

In this case, the steady-state solution satisfies
\begin{equation}
  B_z(x)\left[\frac{J(x)}{J_c}\right]^{n} = \text{const} \quad \Longleftrightarrow \quad B_z^{1/n}(x) \frac{J(x)}{J_c} = \text{const}.
\end{equation}
Since $\frac{J(x)}{J_c} \propto B^{1/n}(x)$, the inward shift of the profile due to the self-field is increasingly suppressed as $n$ grows.
\Cref{fig:n-dep} shows numerical solutions of the $v$--$J$ model as $n$ is varied.
With the present information, we cannot fix the flux-creep exponent $n$ at the relatively high temperature $T \approx \SI{70}{K}$ in our study.
According to \citet{Brandt1996}, a value around $n \approx 10$ is typical in the creep regime.

However, \cref{fig:n-dep} shows that for $n\gtrsim 10$ the computed profile is already more uniform than the experimental one.
This indicates that the vortex distribution is already frozen in a temperature range close to flux flow, and that creep with a large exponent $n\gtrsim 10$ is not reflected.

Indeed, comparing with the resistance versus temperature in the main text, the current distribution shows little temperature dependence once zero resistance is reached, rather than homogenizing as the temperature is lowered. This observation suggests that the vortex distribution becomes fixed in a temperature range very close to the flux-flow regime.

We also show later in \Cref{sec:stab-5mA} that, after sufficient cooling to $72.7$~K, this vortex configuration is very stable against changes in the bias current, which suggests that creep at lower temperatures does not drastically alter the steady-state distribution determined at higher temperature.

\section{Numerical Method}
In this section we describe the numerical procedure used to reconstruct the current distributions in Figs.~3(e) and 5(b) of the main text.
Specifically, starting from a uniformly sampled magnetic-field profile $B_z(x)$, we stably estimate the one-dimensional local current density $J(x)$ inside the strip, $|x|<W$.
As shown in the Appendix A in the main text, the forward and inverse transforms between $J$ and $B$ can be expressed analytically, in a nonsingular form, using Chebyshev polynomials.
Here we formulate the inverse problem, introduce regularization to suppress aliasing, and summarize the relation between distance-induced blurring and the Hilbert transform.

\subsection{Discretization and setup for the transforms}
As in the main text, we choose the sampling points
\begin{equation}
  x_k = W \cos\left(\theta_k\right),\quad
  \theta_k = \frac{2k+1}{2N}\pi,\quad
  k=0,1,\dots,N-1,
  \label{eq:cheb-nodes}
\end{equation}
and henceforth use the dimensionless variable $t=x/W$, so that $\tilde x_k = x_k/W = \cos\theta_k$.
Since the experimental data are given on a uniformly spaced grid $\{B_z(x_j),  x_j = x_0 + \Delta x (j-1)\}$, we first resample them at the locations in \cref{eq:cheb-nodes} by cubic spline interpolation to obtain $\{B^{(k)}_z\}_{k=0}^{N-1}$.

Expanding in Chebyshev polynomials $T_n$, we set
\begin{equation}
  P(t) := J(W t)\sqrt{1-t^2}, \quad
  P(\tilde x_k) = \sum_{n\ge 0} a_n T_n(\tilde x_k)
  = a_0 + \sum_{n\ge 1} a_n \cos(n\theta_k),
  \label{eq:cheb-expand}
\end{equation}
which is consistent with the derivation in the Appendix A in the main text.
Using the parameter $\eta_k$ defined there, the current-to-field transform is
\begin{equation}
  B_z(x_k; d)
  = \frac{\mu_0}{2}\Re \mathcal C[J](x_k + i d)
  = -\frac{\mu_0}{2}\sum_{j\ge 0} a_j\frac{e^{-j\eta_k}}{\sinh \eta_k}
  = \sum_{j\ge 0} K_{kj}a_j,
  \label{eq:forward-K}
\end{equation}
where the last equality defines the matrix $K_{kj}$.

\subsection{Inverse problem and regularization}
\begin{figure}[tbp]
\centering
\includegraphics[width=1.\textwidth]{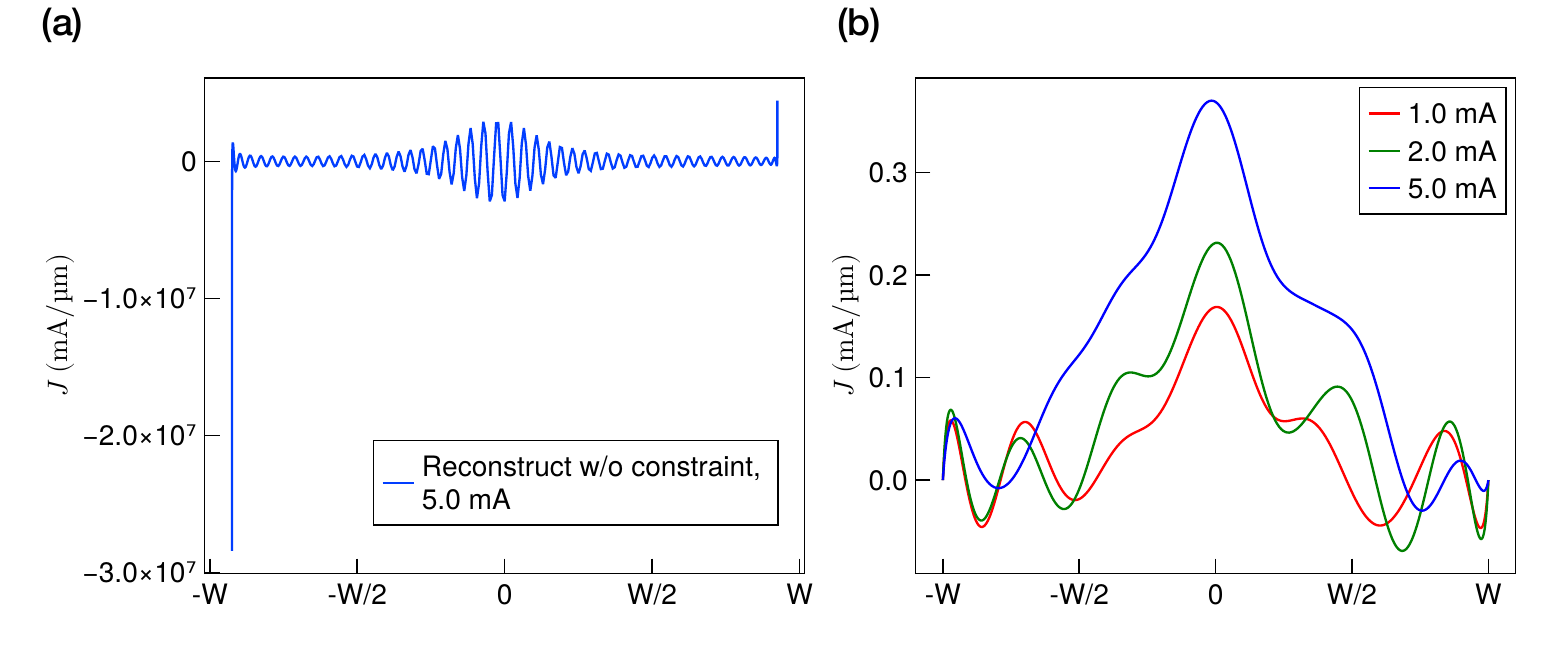}
\caption{
  (a) Result of unconstrained field-to-current reconstruction, showing prominent oscillations.
  (b) Results of minimizing $\mathcal I$ in \cref{eq:cost1} for \SI{1}{mA}, \SI{2}{mA}, and \SI{5}{mA} in Fig.~3(a) of the main text.
}
\label{fig:curr_reconst}
\includegraphics[width=1.\textwidth]{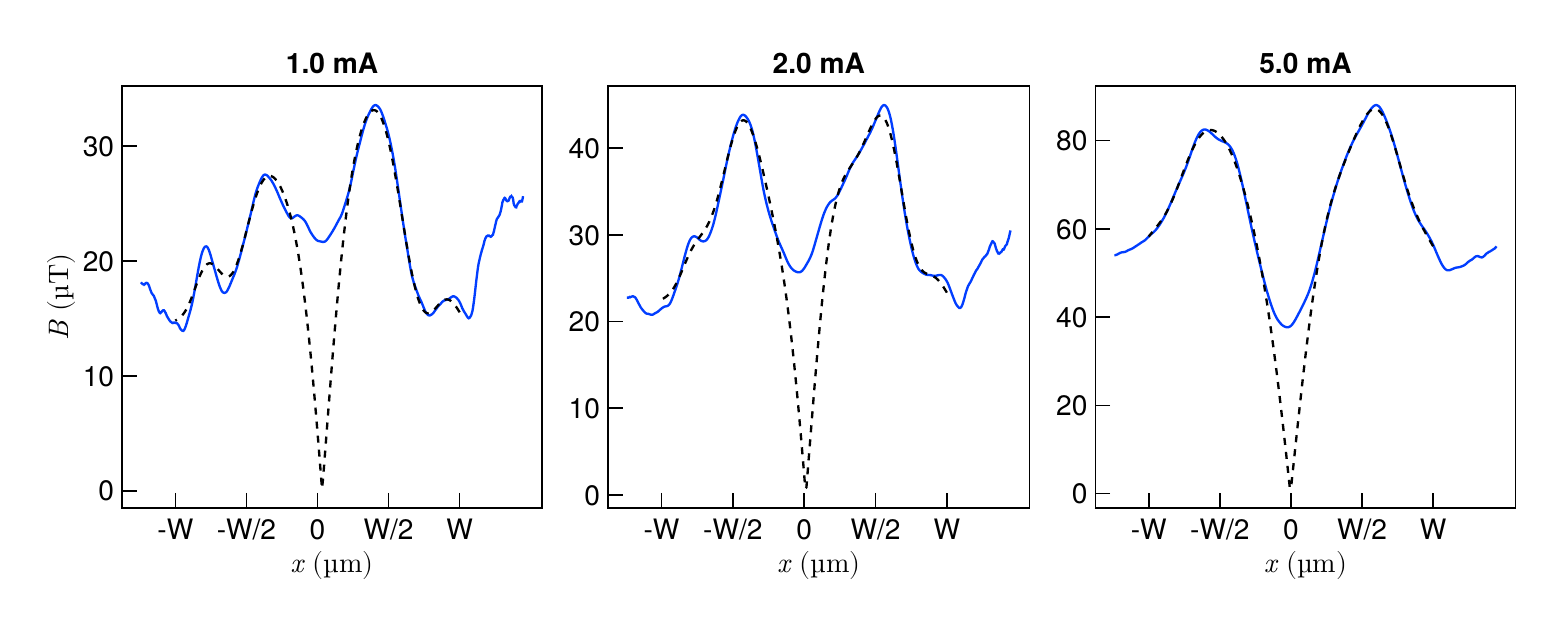}
\caption{
  Comparison between the measured magnetic-field profiles and the fields obtained by transforming the reconstructed currents in \cref{fig:curr_reconst}(b).
  Blue solid lines: experiment. Black dashed lines: reconstruction.
}
\label{fig:field-reconst}
\end{figure}

Because the present NV sensing protocol is sensitive only to the magnitude of the magnetic field, the data must be processed as magnitudes.
If the matrix $K=\{K_{kj}\}$ is invertible, one may in principle apply $K^{-1}$ to a suitably signed magnetic field to obtain $\{a_j\}$.
Alternatively, one can minimize a weighted least-squares cost,
\begin{equation}
  \mathcal I
  = \sum_{k=0}^{N-1}\left(B^{(k)}_z - \Bigl|\sum_{j\ge 0} K_{kj} a_j\Bigr|\right)^{2},
\end{equation}
but, as with Fourier-based approaches, direct inversion encounters aliasing.
\Cref{fig:curr_reconst}(a) shows a fit obtained by na\"\i vely matching the transformed field to the experimental data, where pronounced spurious oscillations appear.

Since the oscillations clearly exceed the true signal, we suppress aliasing while preserving as much field information as possible by adding regularization terms to the cost.

We first penalize current gradients.
Let
\begin{equation}
  P(t) = J(Wt)\sqrt{1-t^2}.
\end{equation}
By adding penalties on $\dv{P}{t}$ and $\dv[2]{P}{t}$,
\begin{equation}
  \Lambda_1 \biggl\| \dv{P}{t} \biggr\|^2 + \Lambda_2 \biggl\| \dv[2]{P}{t} \biggr\|^2,
\end{equation}
we obtain smoother solutions, where $\|\cdot\|$ denotes the $L^2$ norm on $[-1,1]$.
Using
\begin{align}
  \dv{T_n(x)}{x} &= n U_{n-1}(x), \\
  \dv{U_n(x)}{x} &= \frac{(n+1)T_{n+1}(x)-x U_n(x)}{x^{2}-1},
\end{align}
we have
\begin{equation}
  \dv{P}{t} = \sum_{n\geq 1} n a_n U_{n-1}(t),\quad
  \dv[2]{P}{t} = \sum_{n\geq 1} a_n \frac{n^2 T_n(t) - n t U_{n-1}(t)}{t^{2}-1},
\end{equation}
so the derivatives of each polynomial component are available in closed form.

\subsection{Handling sign reversal near the center (weighting)}
As noted in the main text, because the method uses magnitudes, the local field around the strip center yields a biased average and deviates near the sign-reversal point $x=0$.
We therefore relax the fit near $x\approx 0$ by weighting the residuals.
Specifically, we reduce them by a factor of $10^{-2}$ in the region $|x| < \SI{3.0}{\micro m}$.
The final cost function is
\begin{equation}
  \mathcal I = \sum_{k=0}^{N-1} \Lambda_0(x_k)
  \left(B^{(k)}_z - \Bigl|\sum_{j\ge 0} K_{kj} a_j \Bigr|\right)^{2}
  +  \Lambda_1 \biggl\| \dv{P}{t} \biggr\|^2
  + \Lambda_2 \biggl\| \dv[2]{P}{t} \biggr\|^2, \quad
  \Lambda_0(x) = \begin{cases}
    10^{-2} & \text{if } |x| < \SI{3.0}{\micro m},\\
    1 & \text{otherwise.}
  \end{cases}
  \label{eq:cost1}
\end{equation}

\subsection{Implementation settings and validity of the reconstruction}
By minimizing the cost $\mathcal I$, we obtained current distributions for \SI{1}{mA}, \SI{2}{mA}, and \SI{5}{mA} in Fig.~3(a) of the main text, as shown in \cref{fig:curr_reconst}(b).
Here we set $d = \SI{3.0}{\micro m}$, whose validity is discussed later in \Cref{sec:distance}.
Although some oscillations remain, the overall current profiles are preserved.
The magnetic fields computed from these current distributions are shown in \cref{fig:field-reconst}.
Even with the smoothing terms, the field shapes are retained.

\begin{figure}[tbp]
\centering
\includegraphics[width=1.\textwidth]{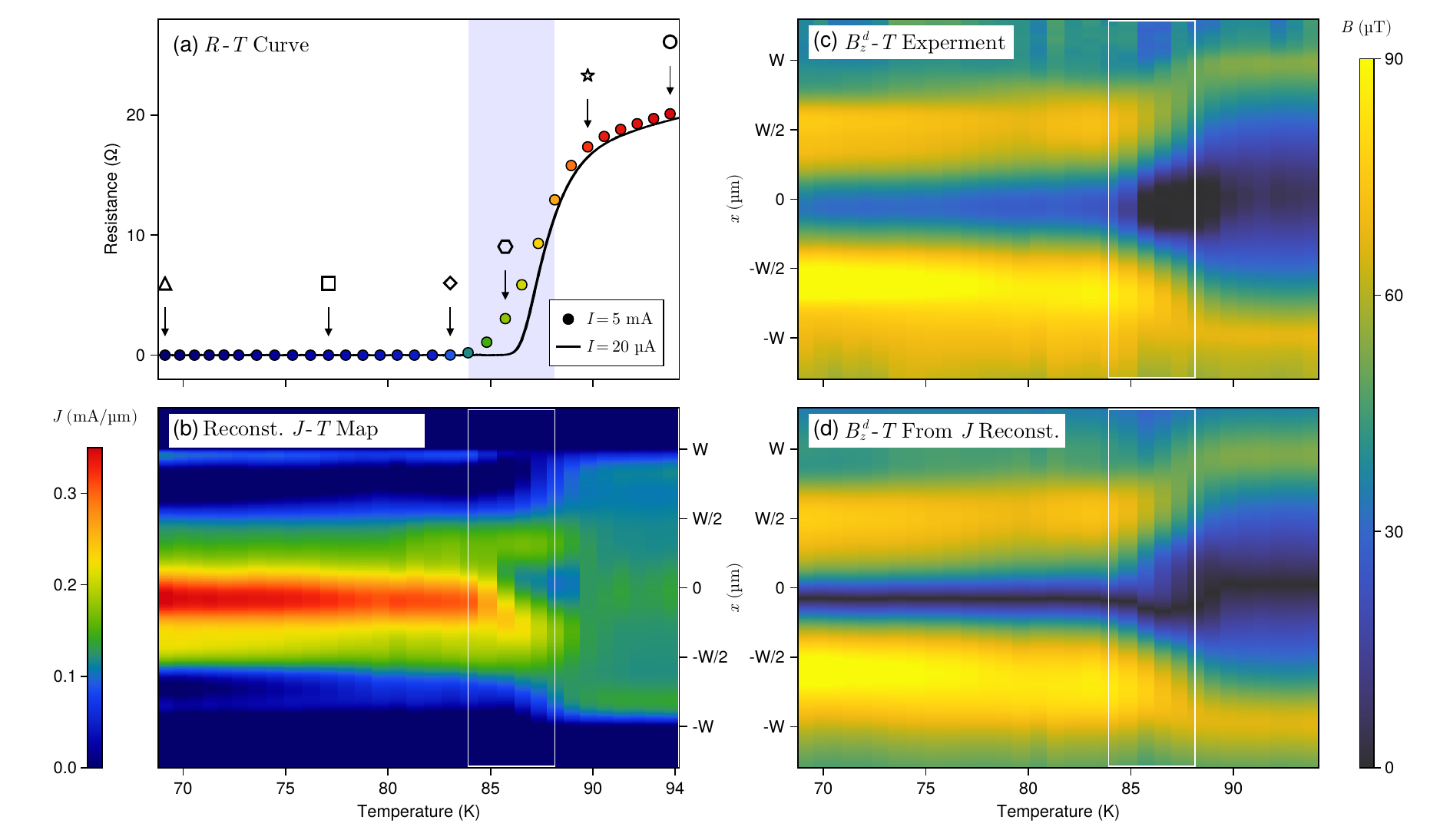}
\caption{
  Comparison between the measured magnetic-field profiles and those obtained from the reconstructed currents.
  (a) Resistance--temperature curve (reproduction of Fig.~5(a) in the main text).
  (b) Two-dimensional intensity map of reconstructed $J$ versus temperature (reproduction of Fig.~5(b) in the main text).
  (c) Two-dimensional intensity map of the experimental field $B_z^{\,d}$ versus temperature.
  (d) Two-dimensional intensity map of the recalculated field $B_z$ from the reconstructed $J$ versus temperature.
}
\label{fig:field-reconst-comp}
\end{figure}

Applying the same procedure, the current distributions in Fig.~5(b) are obtained from the experimental data in Fig.~4(a).
For a temperature-by-temperature comparison, two-dimensional maps of the measured field profiles and those from the reconstructed currents are shown in \cref{fig:field-reconst-comp}.

\subsection{Constraint on the total current and its treatment}
The above calculations do not use the fact that the total current is known.
From the orthogonality of Chebyshev polynomials~\cite{si_TableofIntegrals},
\begin{equation}
  \int_{-1}^{1}T_{n}(t)T_{m}(t)\frac {dt}{\sqrt {1-t^{2}}}=\pi A_{n}\delta _{nm},
  \quad
  A_n =
  \begin{cases}
    1 & \text{if }n = 0,\\
    1/2 & \text{otherwise,}
  \end{cases}
\end{equation}
we obtain for the current integral
\begin{equation}
  I = \int_{-W}^{W} J(x)  dx
  = W \sum_{n\geq 0}\int_{-1}^1 \frac{a_n T_n(t)}{\sqrt{1-t^2}}  dt
  = \pi a_0 W .
\end{equation}
Thus, by fixing the first coefficient, one can enforce exact agreement of the total current with the experimental value during optimization.

On the other hand, enforcing a fixed current introduces aliasing near the strip ends, which we attribute to the distance ambiguity discussed below.
Therefore, in \cref{fig:curr_reconst} we select $d$ so that the reconstructed current does not deviate significantly from the actual value and optimize without constraining the total current.
With $d = \SI{3.0}{\micro m}$, the integrated currents are \SI{1.34}{mA}, \SI{2.07}{mA}, and \SI{4.95}{mA} for \SI{1.0}{mA}, \SI{2.0}{mA}, and \SI{5.0}{mA}, respectively.
Although the error is larger at low current, the effect of not fixing the current diminishes as the current increases.
Moreover, the choice $d = \SI{3.0}{\micro m}$ is reasonable on average, given the sample-to-NV distance $\SI{1.35}{\micro m}$ and the NV-layer thickness $\SI{2.3}{\micro m}$ reported in the prior work~\cite{si_Nishimura2023}.

\subsection{Effects of the distance between the sensor layer and the sample}
\label{sec:distance}
Our measurements are affected by blurring from the sensor–sample separation, optical resolution, and the finite thickness of the sensor layer.
These effects cannot be separated: the magnetic field at a distance $d$ is
\begin{equation}
\Bzd(x)=\frac{\mu_0}{2\pi}\int_{-W}^{W}\frac{u-x}{(u-x)^2+d^2}J(u)du\quad\text{for }x\in \mathbb R,
\end{equation}
which, upon extending $J$ to $\mathbb R$ with $J(x)=0$ for $|x|>W$, can be written as a convolution with the kernel
\begin{equation}
  Q_d(s) = \frac{s}{\pi (s^2 + d^2)}, \qquad
  \Bzd(x)=\frac{\mu_0}{2} (Q_d * J)(x).
\end{equation}
The kernel $Q_d$ is the harmonic conjugate of the Poisson kernel
\begin{equation}
  P_d(s) = \frac{d}{\pi (s^2 + d^2)},
\end{equation}
and satisfies
\begin{equation}
  Q_d = h * P_d, \qquad h(s)=\frac{1}{\pi s},
\end{equation}
where $h$ is the impulse response of the Hilbert transform,
\begin{equation}
  \mathcal{H}[f](x) = \mathrm{p.v.} 
  \int_{\mathbb R} 
  \frac{f(u)}{\pi(u-x)}  du = (h * f)(x).
\end{equation}
Hence,
\begin{equation}
  \Bzd(x) = \frac{\mu_0}{2} [h * (P_d * J)](x) = \frac{\mu_0}{2} \mathcal H \left[P_d * J\right](x).
\end{equation}
Thus, measuring at a distance $d$ is equivalent to blurring by the Poisson kernel $P_d$ followed by a Hilbert transform.
Any additional blur of the magnetic field, due to factors besides distance, appears as an ambiguity in $d$.
In terms of the total current, such blur can bias the estimate.
Underestimating $d$ makes the forward kernel stronger, hence the inversion tends to underestimate $J$ locally, while overestimating $d$ has the opposite effect.
This issue, however, does not strongly influence the overall current shape and therefore does not significantly affect the present discussion.

\section{Experimental Methods and Conditions}
\begin{figure}[tbp]
\centering
\includegraphics[width=1.\textwidth]{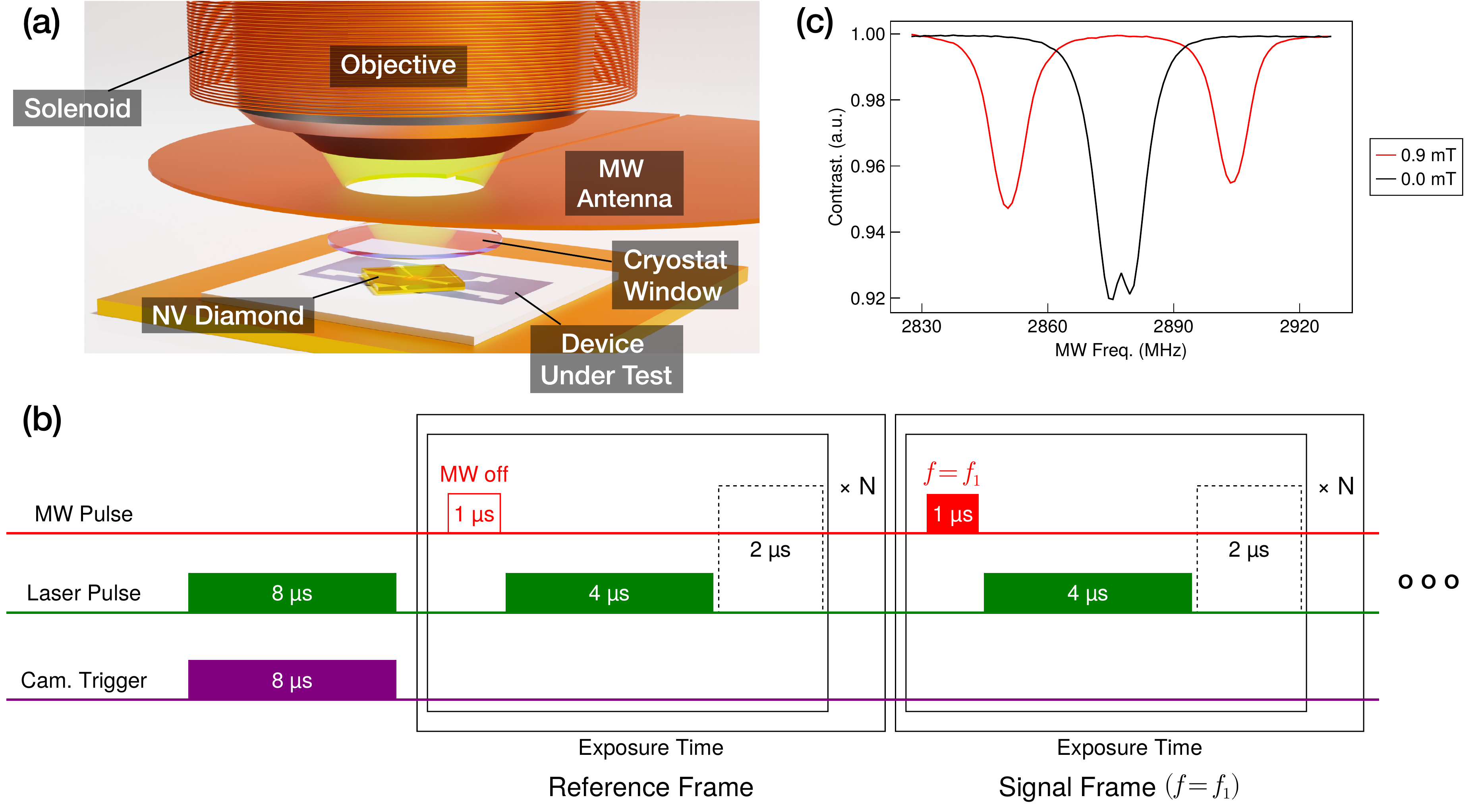}
\caption{
  (a) Schematic of our microscope.
  (b) Typical ODMR spectrum obtained using the pulse protocol in (c).
  (c) Diagram of the pulsed-ODMR (p-ODMR) protocol. Because the image sensor operates far more slowly than the time scales of NV initialization and control, we repeat the same sequence during the exposure time and read out the accumulated signal. We then sweep the microwave frequency and repeat the protocol at each point to obtain the ODMR spectrum.
}
\label{fig:setups}
\end{figure}

\subsection{Details of Magnetic-Field Measurements Using NV Centers}

In this study we use a home-made quantum diamond microscope (QDM) based on an ensemble of nitrogen-vacancy (NV) centers~\cite{si_Taylor2008,si_Levine2019,si_Nishimura2023} [\cref{fig:setups}(a)].
As the sensor layer, we employ a perfectly oriented NV layer~\cite{si_Nishimura2023,si_Tsuji2022} grown to a thickness of approximately \SI{2.4}{\micro\meter} on the diamond surface, enabling imaging of the magnetic flux density perpendicular to the superconducting thin film.
The magnetic flux density is obtained via optically detected magnetic resonance (ODMR) of the NV centers~\cite{si_Rondin2014}. \Cref{fig:setups}(b) shows a typical ODMR spectrum. The vertical axis is the photoluminescence (PL) intensity normalized to the presence or absence of microwave irradiation, and the horizontal axis is the microwave frequency. Each spectrum exhibits two dips corresponding to the electronic spin resonances of the NV centers. The splitting $\Delta f$ between the resonance frequencies reflects the Zeeman effect and is larger at \SI{0.9}{\milli\tesla} (red squares) than at \SI{0}{\milli\tesla} (black circles). The splitting is given by~\cite{si_Rondin2014}
\begin{equation}
\Delta f = 2\sqrt{ (\gamma_{e}B_z)^2 + E^2},
\label{eq1}
\end{equation}
where $B_z$ is the magnetic flux density along the NV axis, $\gamma_e=\SI{28}{\mega\hertz\per\milli\tesla}$ is the electron-spin gyromagnetic ratio, and $E$ is a strain parameter that varies with position in the crystal. At each pixel we fit the ODMR spectrum to the sum of two Lorentzian functions to extract $\Delta f$.
We also acquire a zero-field reference measurement and identify $\Delta f(B=0)$ with $E$. Using this zero-field reference and \cref{eq1}, we convert $\Delta f$ to $B_z$. The use of perfectly oriented NV centers simplifies the analysis: a general multi-orientation ensemble can exhibit up to eight resonance lines, which complicates fitting.
The absence of NV orientations aligned to other symmetry axes avoids contrast loss, improving sensitivity. The magnetic-field map is obtained by analyzing all \num{2048}\,$\times$\,\num{2048} CMOS pixels. To suppress failures in the Lorentzian fitting, we smooth the PL images with a Gaussian filter smaller than the optical resolution to reduce shot noise (the filter's $1/e$ decay length is \SI{350}{\nano\meter}, corresponding to five camera pixels).

\subsection{Microscope Configuration}

\Cref{fig:setups}(a) shows a schematic of the measurement setup. The sample temperature is controlled by a heater on the cold stage and monitored with a thermometer attached to the stage. In addition, a solenoid coil applies a static magnetic field perpendicular to the YBCO surface. For PL imaging of the NV centers, we excite the diamond with a green laser (\SI{515}{\nano\meter}, \SI{140}{\milli\watt}) and collect the NV PL using an image sensor (Andor Zyla ZL41 4.2) placed outside the cryostat optical window. The microwave antenna required for NV manipulation (a loop-type resonator~\cite{si_Sasaki2016}) and the solenoid are mounted outside the optical cryostat to avoid heating.

\begin{figure}[tbp]
\begin{center}
\includegraphics[width =1.\textwidth]{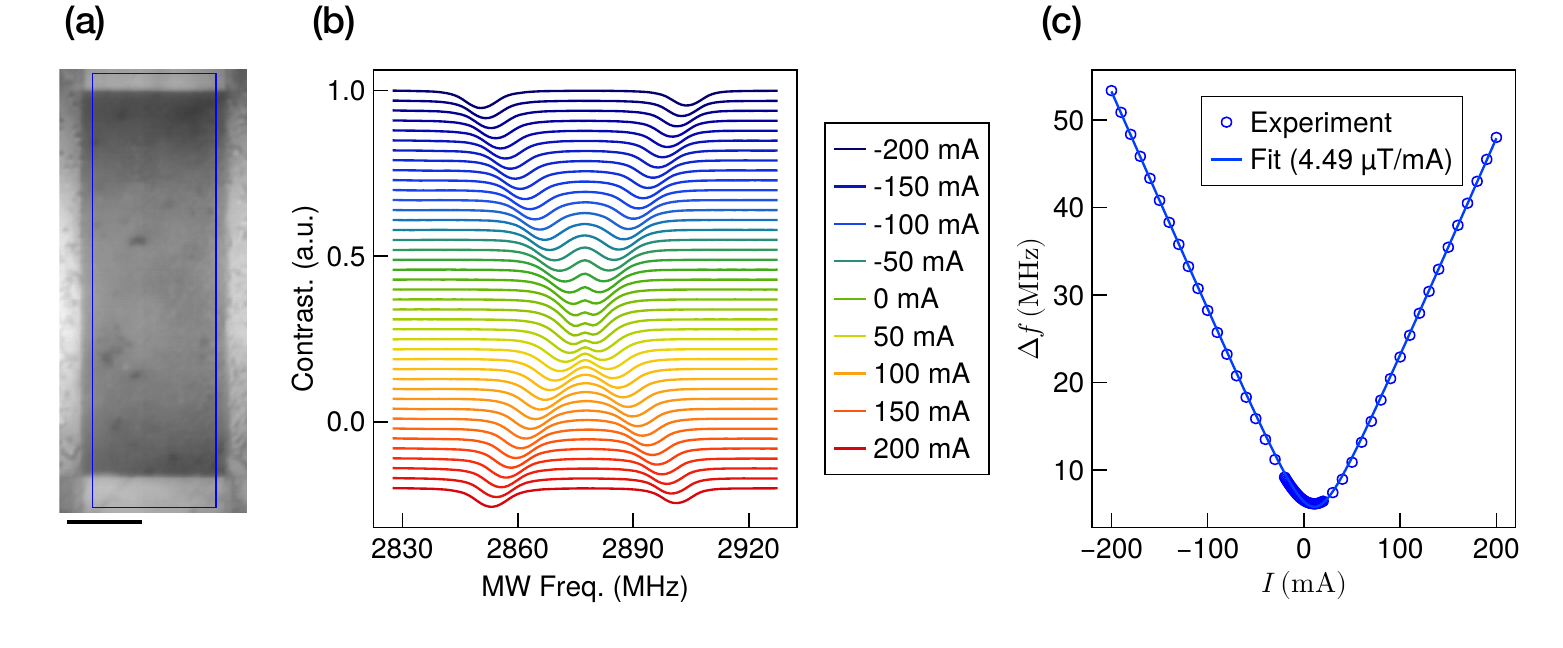}
\caption{
  Calibration experiment. 
  (a) Bright-field image of the device captured by the QDM microscope. The scale bar corresponds to \SI{20}{\micro m}.
  (b) Dependence of the resonance-frequency difference $\Delta f$ on the current applied to the solenoid.
  We average $\Delta f$ over the calibration region indicated by the blue rectangle in \cref{fig:20ref}(a).
  We use this result to calibrate the uniform external magnetic-field bias.
}
\label{fig:20ref}
\end{center}
\end{figure}

\subsection{Pulse Protocol}
We perform the NV magnetic-field measurements using the pulse protocol shown in \cref{fig:setups}(c). The time-averaged laser input power is therefore approximately $\SI{140}{mW} \times \frac{4}{4 + 2 + 1} \approx \SI{80}{mW}$. Waveforms are generated with an arbitrary waveform generator (Spectrum M4.6631-x8). The laser is the dominant heat source, which locally heats the sample and causes the actual sample temperature to deviate from the thermometer reading at the temperature-controlled stage. Therefore, except during transport measurements, we keep the laser pulse waveform running continuously while acquiring data. We calibrate this temperature deviation by performing current--voltage (I--V) measurements on the same sample with the laser toggled on and off; details are provided in \cref{subsec:laseron}. Because the pulse-cycle period is very short (\SI{7}{\micro\second}), thermal inhomogeneities cannot follow the cycle, so the temperature rises quasi-steadily in a time-averaged sense.

\subsection{Magnetic-Field Calibration}

We install a solenoid coil to apply finely controlled magnetic fields to the superconductor, enabling modulation with a precision of $\lesssim \SI{5}{\micro\tesla}$. This uncertainty arises from fluctuations of the ambient field, including the geomagnetic field, rather than from the reproducibility of the field generated by the coil.
By adjusting the current applied to the coil, we set the magnetic field, and we add a finite offset current to compensate the geomagnetic field. Thus, the coil current $I_\text{coil}$ and the resulting $z$-directed magnetic field $B_{\mathrm{ext},z}$ satisfy
\begin{equation}
B_{\mathrm{ext},z} = \alpha \bigl(I_\text{coil} - I_0\bigr),
\label{eq:calib1}
\end{equation}
where the zero-field offset current $I_0$ (mA) and the current-to-field conversion factor $\alpha$ (\si{\micro\tesla\per\milli\ampere}) are determined by applying relatively strong fields and using the NV relation below.

Because the NV-center frequency splitting $\Delta f$ depends nonlinearly on the field magnitude near zero field,
\begin{equation}
\Delta f = 2\sqrt{(\gamma_e B)^2 + E^2},
\label{eq:calib2}
\end{equation}
fitting the data with \cref{eq:calib1,eq:calib2} yields
\begin{equation}
\alpha = 4.49(1)~\si{\micro\tesla\per\milli\ampere}, \quad I_0 = 10.7(5)~\si{mA}.
\end{equation}

\begin{figure}[tbp]
\centering
\includegraphics[width=0.7\textwidth]{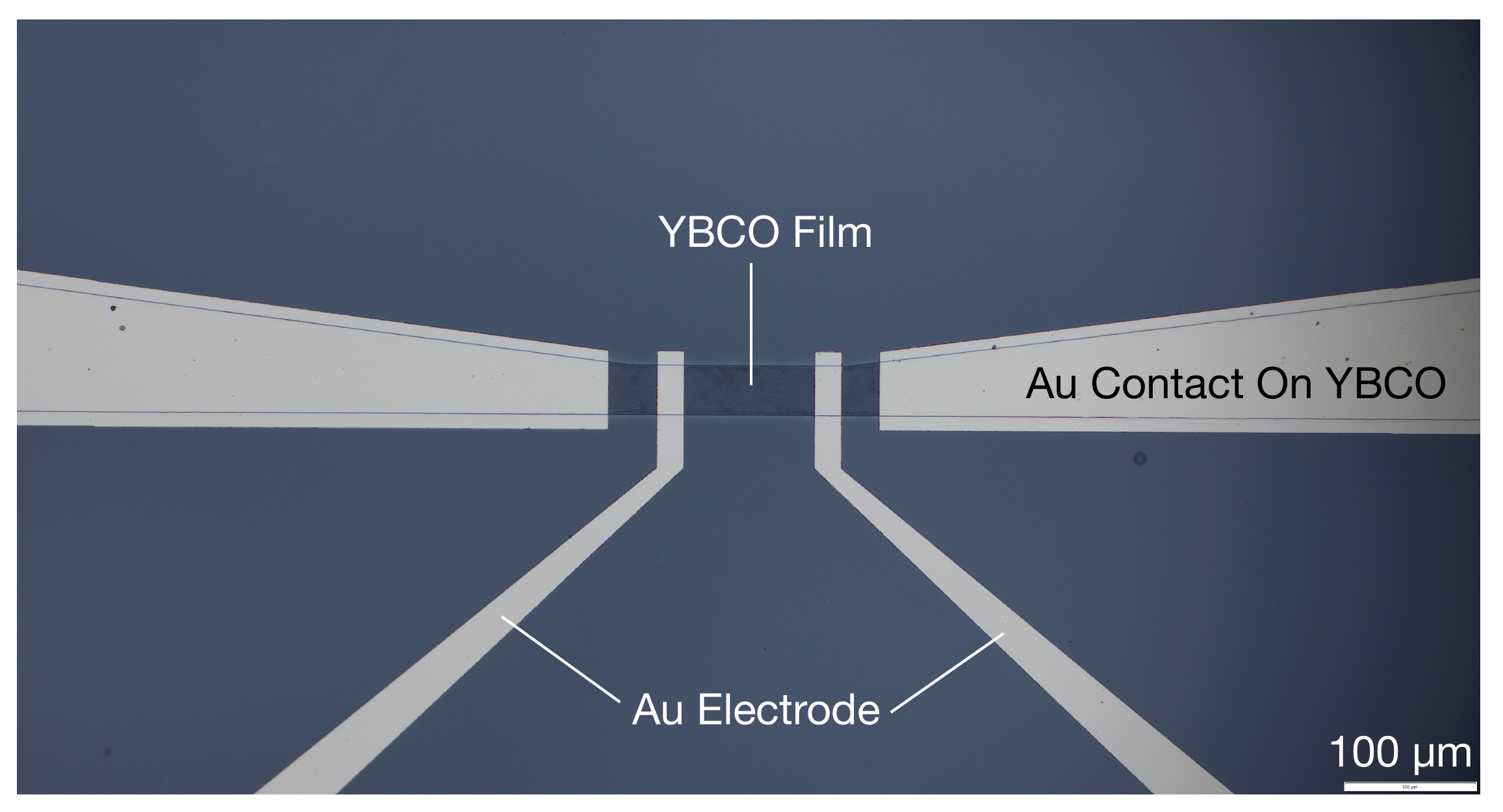}
\caption{
Optical micrograph of the YBCO microstrip.
}
\end{figure}

\begin{figure}[tbp]
\centering
\includegraphics[width=1.\textwidth]{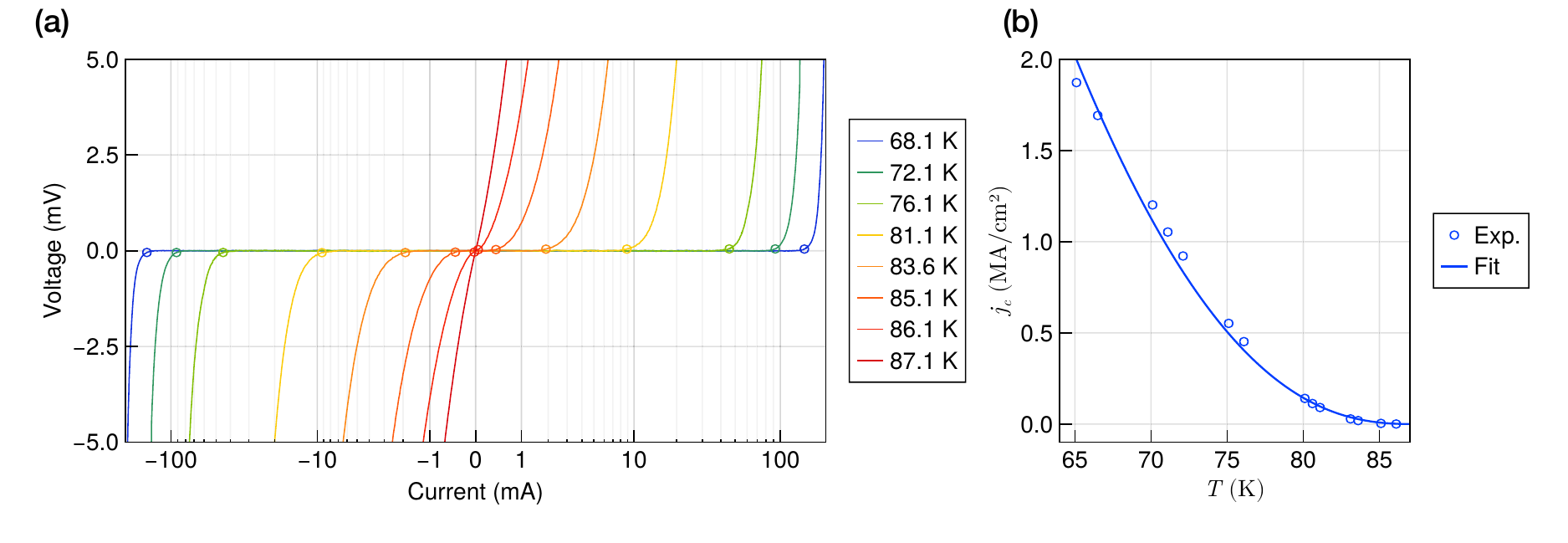}
\caption{
(a) Temperature-dependent I--V curves of the YBCO thin-film strip.
(b) Temperature dependence of the critical current density obtained from (a).
The critical current density is determined as the offset current threshold at which the voltage exceeds $3\sigma$, where $\sigma$ is the voltage fluctuation in the zero-resistance regime.
For the temperature calibration described below, the solid line is a cubic-spline interpolation of the $j_c(T)$ curve.
}
\label{fig:IV}
\end{figure}

\begin{figure}[tbp]
\centering
\includegraphics[width=1.\textwidth]{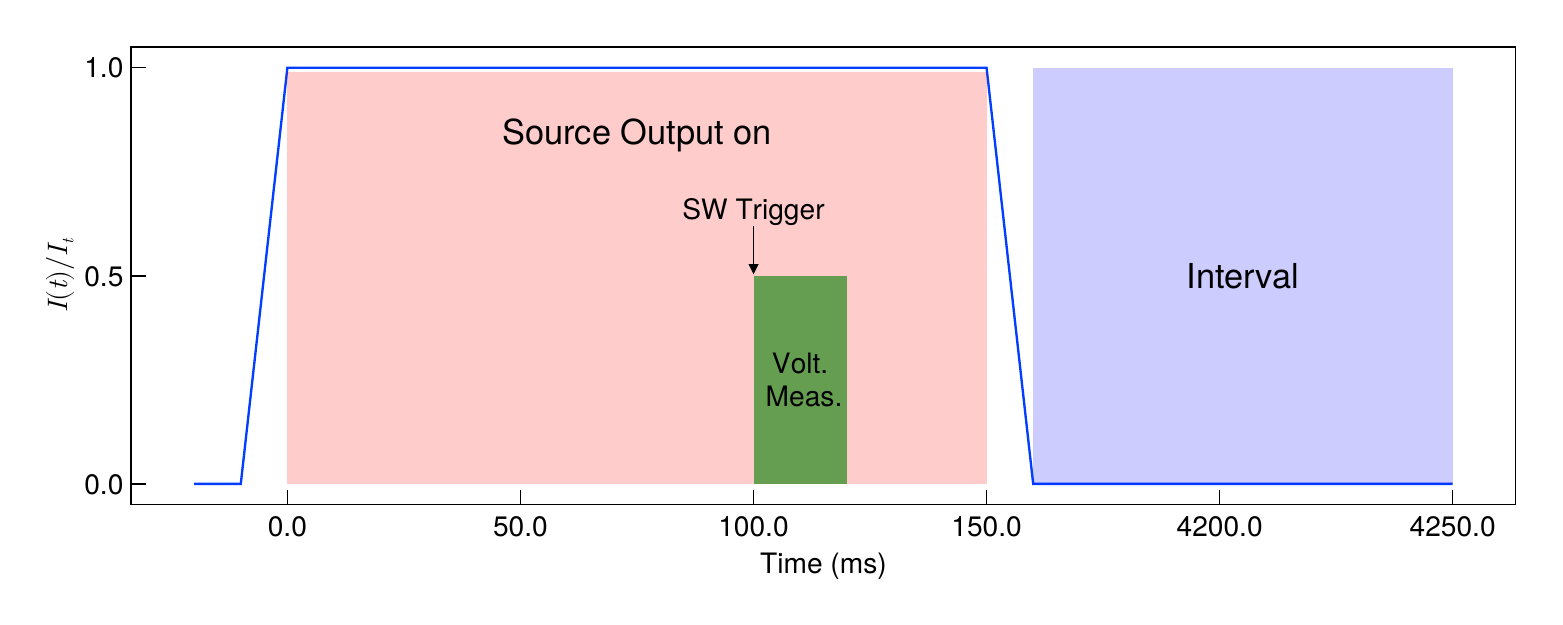}
\caption{
Measurement scheme for DC pulses in the I--V measurement. The current is applied for a total of about \SI{150}{ms} to stabilize, and the voltage is measured starting at \SI{100}{ms}.
With the power-line cycle set to 1, the voltage is integrated over exactly \SI{20}{ms}. Other timings are software controlled and may carry small errors.
We repeat the measurement 20 times and average to improve the signal-to-noise ratio and to determine the critical current accurately.
}
\label{fig:IV-pulse}
\end{figure}

\subsection{YBCO Sample}

We use a commercial \SI{250}{\nano\meter}-thick YBCO film purchased from Ceraco. The film is patterned by argon ion milling. The strip width is set to \SI{37}{\micro\meter}, chosen so that individual vortices appear optically discrete over our measurement range. Four \SI{50}{\nano\meter}-thick gold electrodes are deposited as voltage terminals and positioned so the terminals lie within the field of view. To ensure good contact between YBCO and the gold electrodes, we first perform Ar-ion milling the YBCO surface by a few nanometers, then evaporate \SI{1}{\nano\meter} of Ti followed by Au, both steps by electron-beam evaporation. We also enlarge the contact overlap to increase the contact area. These procedures yield metallic contacts, confirmed by the temperature dependence of the two-terminal resistance measured between the current leads.

I--V measurements of this device taken in the same cryostat are shown in \cref{fig:IV}(a).
The superconducting transition temperature is estimated to be \SI{86.1}{K}.
The temperature dependence of the critical current in the unit of $j$ is shown in \cref{fig:IV}(b).
The $j_c(T)$ curve is well reproduced by the mean-free-path fluctuation model~\citet{Griessen1994}, as fitted by the blue solid line in \cref{fig:IV}(b).

For these critical-current measurements we use a Yokogawa GS200 as the current source and a Keithley 2182A for voltage readout. Since up to \SI{200}{mA} DC is applied, heating must be suppressed. For the I--V measurements only, we therefore apply DC pulses as shown in \cref{fig:IV-pulse}.

\subsection{Temperature Calibration}\label{subsec:laseron}

\begin{figure}[tbp]
\centering
\includegraphics[width=1.\textwidth]{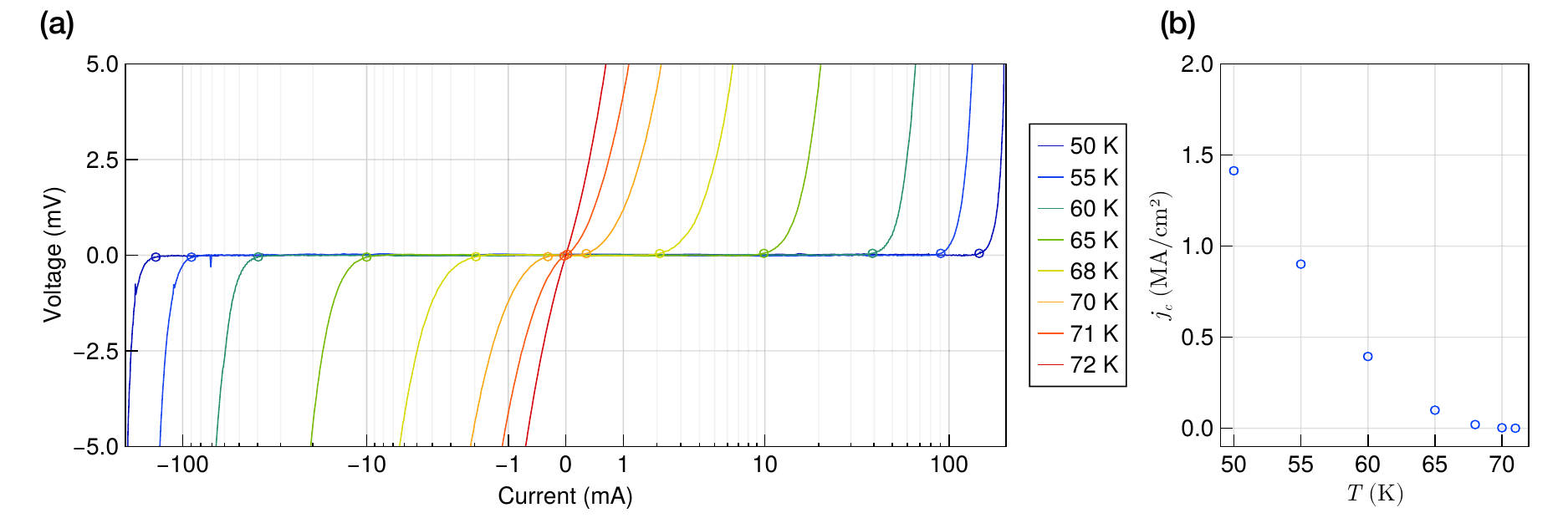}
\caption{
(a) Temperature-dependent I--V curves of the YBCO thin-film strip during laser irradiation with the pulse cycle in \cref{fig:setups}(c).
(b) Temperature dependence of the critical current density obtained from (a).
}
\label{fig:IV-laseron}
\end{figure}

\begin{figure}[tbp]
\centering
\includegraphics[width=1.\textwidth]{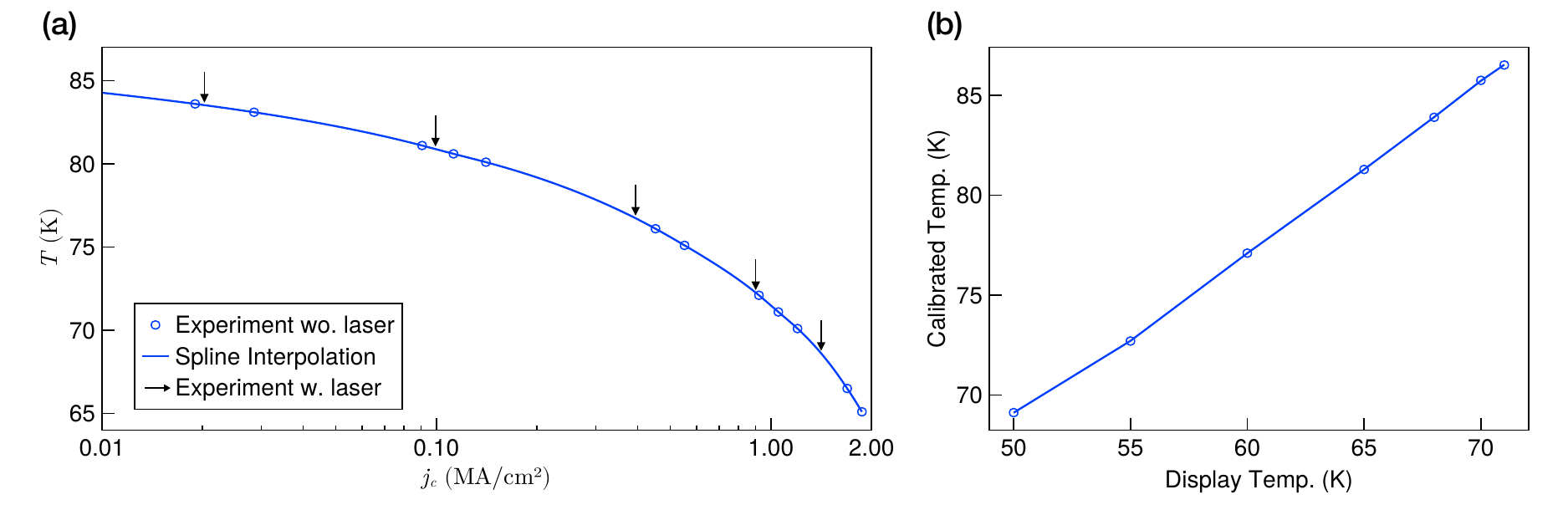}
\caption{
  (a) Spline interpolation of the data points in \cref{fig:IV}(b). The $x$ coordinates of the black arrows correspond to the data points in \cref{fig:IV-laseron}(b), and the $y$ coordinates of the black arrows are the estimated temperatures.
  (b) The temperature reconstructed from $j_c$ (Calibrated Temp.) as a function of the platform's indicated temperature (Display Temp.), based on the spline interpolation of \cref{fig:IV}(b).
}
\label{fig:T-recover}
\end{figure}

We calibrate the temperature deviation induced by the laser pulse waveform of \cref{fig:setups}(c) by performing I--V measurements on the same sample with the laser toggled on and off.
\Cref{fig:IV-laseron} shows the results. The temperature indicated is that of the thermometer mounted on the platform of the cold stage. Largely independent of the platform temperature, the sample temperature increases by approximately \SI{15}{K} as an almost constant offset. The resulting $j_c(T)$ curve is shown in \cref{fig:IV-laseron}(b). Using each point as a sample, we reconstruct the sample temperature from the measured $j_c$ by referencing the spline interpolation of \cref{fig:IV}(b), which is shown in \cref{fig:T-recover}(a). The reconstructed relation is plotted in \cref{fig:T-recover}(b). Temperatures quoted in the main text are obtained by linear interpolation or extrapolation of this curve.

\begin{figure}[tbp]
\centering
\includegraphics[width=0.7\textwidth]{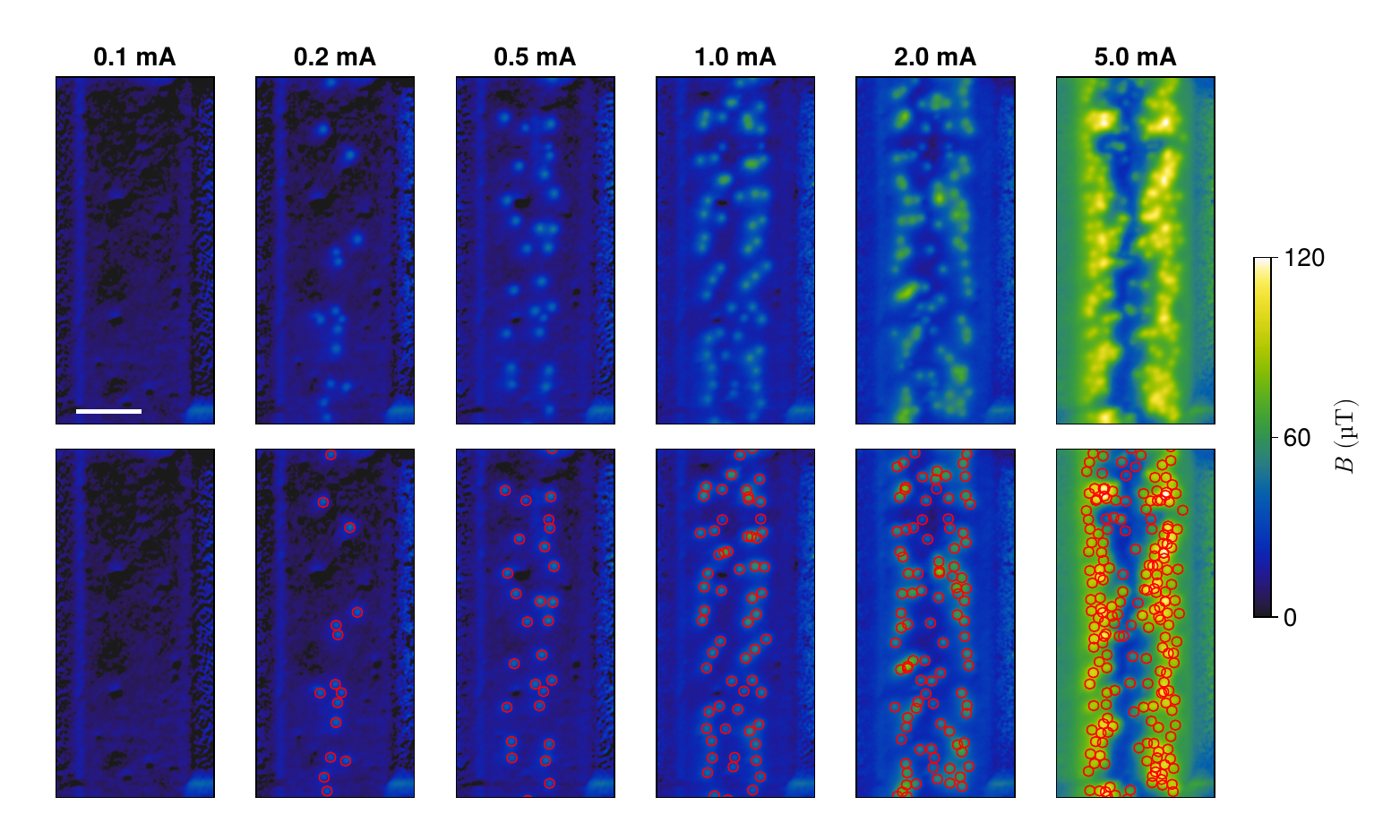}
\caption{
  (Upper row) Magnetic-field maps of the extended datasets used for Fig.~3(a) in zero applied field and for Figs.~3(c,d) in the main text.
  (Lower row) Points from the assigned and extracted peaks (red scatter plots) overlaid on each dataset in the upper row.
}
\label{fig:CC-ZF}
\centering
\includegraphics[width=0.7\textwidth]{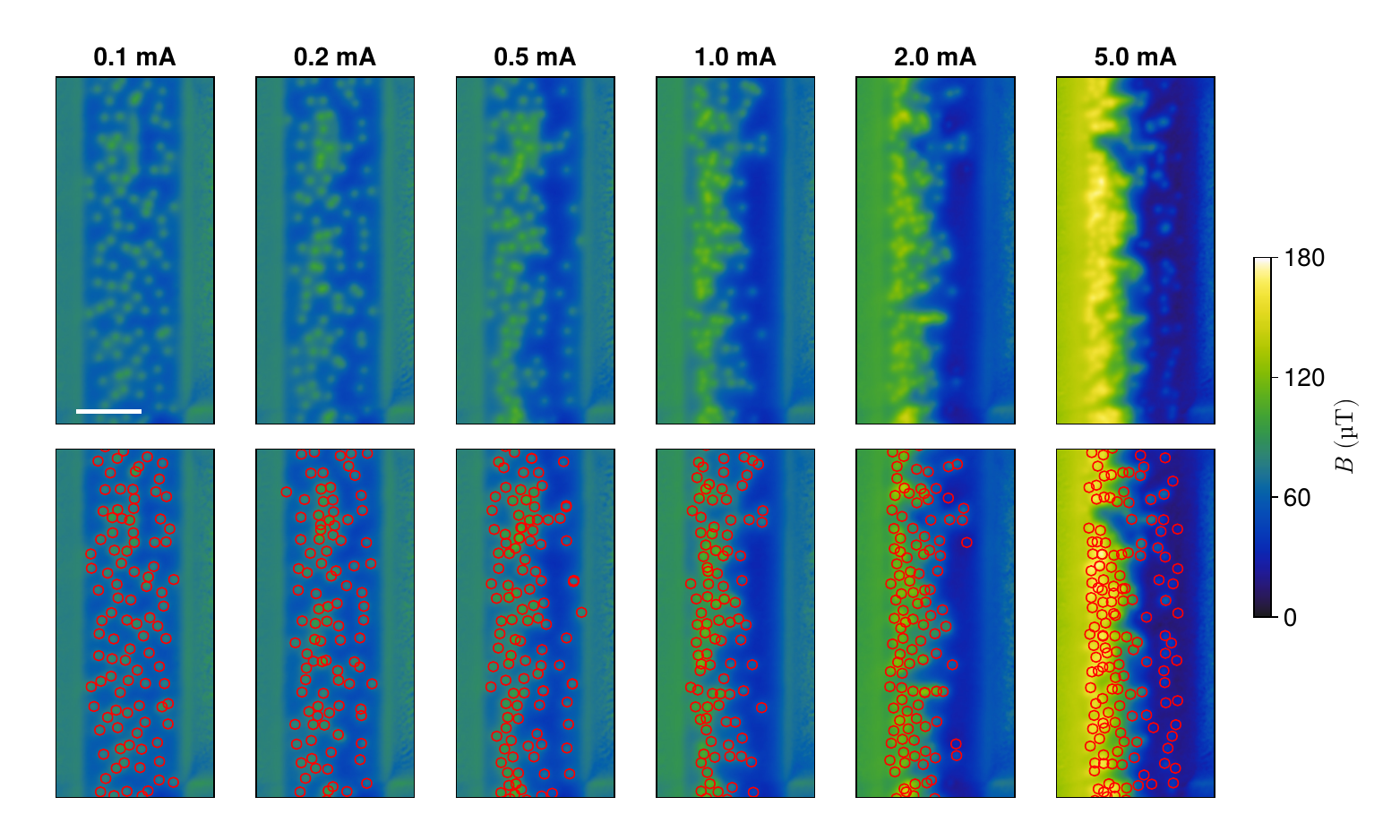}
\caption{
  (Upper row) Magnetic-field maps of the extended datasets used for Fig.~3(b) in the main text (applied field \SI{86.4}{\micro\tesla}) and for Figs.~3(c,d).
  (Lower row) Points from the assigned and extracted peaks (red scatter plots) overlaid on each dataset in the upper row.
}
\label{fig:CC-FF}
\end{figure}

\begin{figure}[tbp]
\centering
\includegraphics[width=1.\textwidth]{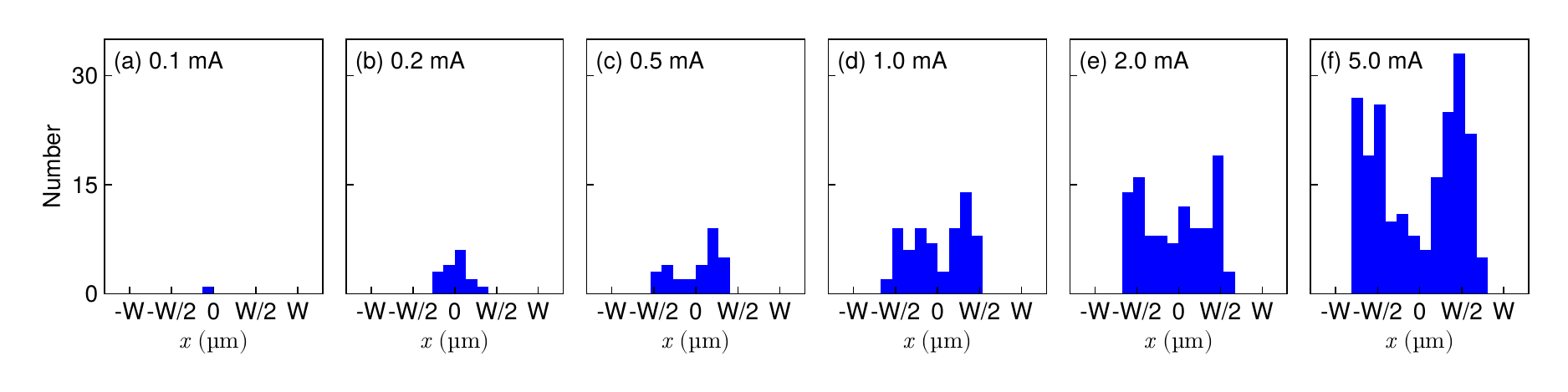}
\caption{
  Histogram of the $x$ coordinates of the assigned points in the lower row of \cref{fig:CC-ZF}.
}
\label{fig:CC-ZF-hist}
\end{figure}

\begin{figure}[tbp]
\centering
\includegraphics[width=1.\textwidth]{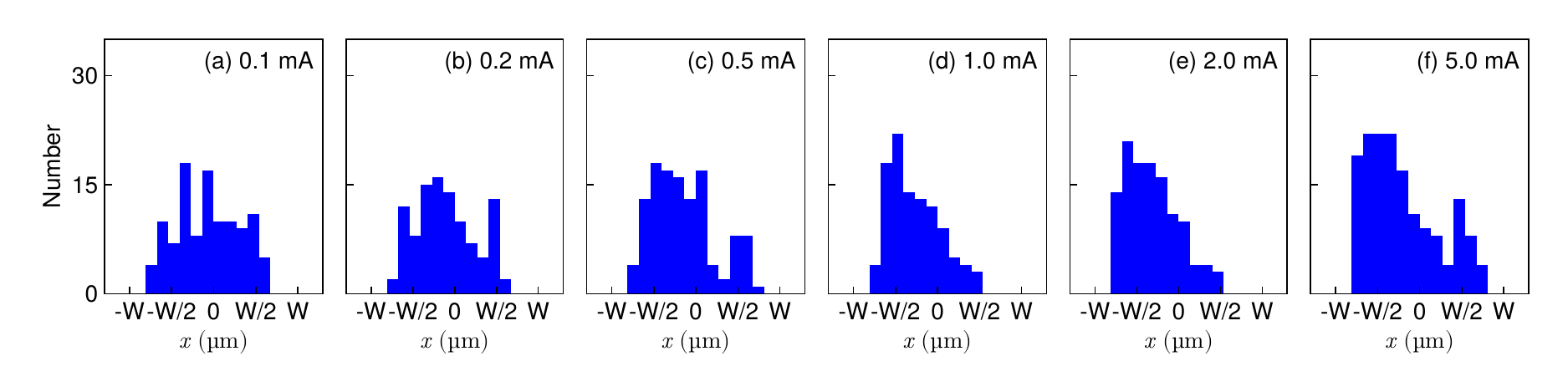}
\caption{
  Histogram of the $x$ coordinates of the assigned points in the lower row of \cref{fig:CC-FF}.
}
\label{fig:CC-FF-hist}
\end{figure}

\newpage

\section{Supplementary Data}
In this section we present supplementary data that are not shown in the main text.

\subsection{Vortex position assignment data}
First, in \cref{fig:CC-ZF} we present the extended set of current-dependence data obtained by current-biased cooling for Fig.~3(a) of the main text (zero applied magnetic field) that allows peak extraction ($< \SI{10}{mA}$).
In the lower row, for each dataset, we overlay the points obtained by assigning and extracting peaks (red scatter plots).
Based on this peak extraction, we obtained the count scaling shown in Fig.~3(c).

Next, \cref{fig:CC-ZF-hist} shows the histogram of the $x$ coordinates of the vortex positions obtained from this peak extraction.
The result in Fig.~3(d) is consistent with this histogram.

Likewise, in \cref{fig:CC-FF} we present the extended set of current-dependence ($< \SI{10}{mA}$) data obtained by current-biased cooling for Fig.~3(b) of the main text (applied field \SI{86.4}{\micro\tesla}).
The lower row overlays, on each dataset, the points obtained by assigning and extracting peaks (red scatter plots).
The histogram of the $x$ coordinates of the vortex positions obtained from this peak extraction is shown in \cref{fig:CC-FF-hist}.

\subsection{Stability of the vortex configuration after \SI{5}{mA} current-biased cooling}
\label{sec:stab-5mA}
\begin{figure}[bht]
\centering
\includegraphics[width=0.7\textwidth]{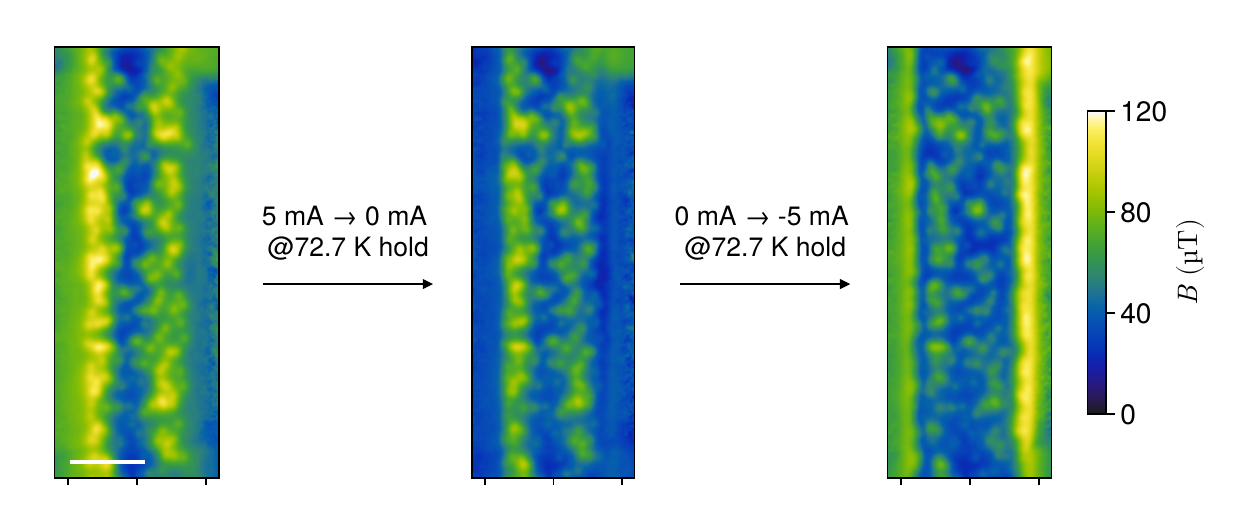}
\caption{
  (Left) Magnetic-field distribution after current-biased cooling at \SI{5}{mA} (down to \SI{72.7}{K}).
  (Center) Magnetic-field distribution measured immediately after the left panel, with the current changed to \SI{0}{mA}.
  (Right) Magnetic-field distribution measured immediately after the center panel, with the current changed to \SI{-5}{mA}.
}
\label{fig:5mA-stability}
\end{figure}

In the main text we noted that, after cooling, vortices freeze and do not move.
Here we show experimentally that the vortex distribution remains highly stable even when the current is changed after current-biased cooling.
\Cref{fig:5mA-stability} shows magnetic-field distributions measured after current-biased cooling to \SI{72.7}{K}, followed by decreasing and reversing the current.
From \cref{fig:5mA-stability} (center), one sees that the vortex positions do not change, whereas the background field does.
This indicates that, in addition to determining the vortex distribution, the transport current biases the overall magnetic-field distribution.
Furthermore, in the result with the current reversed (rightmost panel), the field distribution localized at the strip edges and the vortex distribution appear superposed.
This behavior is well understood by Bean's critical-state model for transport current~\cite{si_Brandt1993,si_Zeldov1994Magnetization}, once the vortex positions inside the strip are taken to be robustly fixed and the rest of the strip is regarded as being in the Meissner state.
From these observations, the vortex positions are frozen robustly enough to withstand perturbations comparable to the magnitude of the present applied current.

